\documentclass[useAMS,usenatbib]{mn2e}
\usepackage{graphicx,amssymb, txfonts}
\usepackage{caption}
\usepackage{footnote}
\usepackage{tablefootnote}
\usepackage{float}

\def\h2{H$_2$}
\def\p1{Paper~I}

\def\kms{$\rm km\,s^{-1}$}
\def\str{{\sc starlight}}

\usepackage{color}

\title[Mapping the inner kpc of NGC~5044 ]{Integral Field Spectroscopy of the inner kpc of the elliptical galaxy NGC~5044}
\author[Diniz et al.]{Suzi I.F. Diniz$^{1}$\thanks{E-mail:
suzi.diniz@ufrgs.br}, Miriani G. Pastoriza$^{1}$, Jose A. Hernandez-Jimenez$^{1,2}$, Rogerio Riffel$^{1}$, 
\newauthor  Tiago V. Ricci$^{2}$, Jo\~ao E. Steiner$^{3}$, Rogemar A. Riffel$^{4}$ \\
1. Universidade Federal do Rio Grande do Sul, Instituto de F\'\i sica, CP 15051, 91501-970, Porto Alegre, RS, Brazil.\\
2. Universidade Federal da Fronteira Sul, Campus Cerro Largo, Rua Major Antonio Cardoso, Cerro Largo, RS, 97900-000, Brazil.\\
3. Instituto de Astronomia, Geof\'\i sica e Ci\^encias Atmosf\'ericas, Universidade de S\~ao Paulo, 05508-900, S\~ao Paulo, Brasil\\
4. Universidade Federal de Santa Maria, Departamento de F\'\i sica, Centro de Ci\^encias Naturais e Exatas, 97105-900, Santa Maria, RS, Brazil.\\
}
\begin{document}

\date{}

\pagerange{\pageref{firstpage}--\pageref{lastpage}} \pubyear{2016}

\maketitle

\label{firstpage}

\begin{abstract}

{ We used Gemini Multi-Object Spectrograph (GMOS) in the Integral Field Unit 
mode to map the stellar population, emission line flux distributions and gas
kinematics in the inner kpc of NGC~5044. From the stellar populations synthesis 
we found that the continuum emission is dominated by old high metallicity stars ($\sim$13~Gyr, 2.5\,Z$\odot$). Also, its nuclear emission is diluted by a non 
thermal emission, which we attribute to the presence of a weak active galactic nuclei (AGN). In addition, 
we report for the first time a broad component (FWHM$\sim$ 3000 \kms) in the H$\alpha$ emission line in the nuclear region of NGC\,5044. 
By using emission line ratio diagnostic diagrams we found that two dominant ionization processes coexist, while the nuclear region (inner 200\,pc) is ionized by a low luminosity AGN,  { the filamentary structures are consistent with being excited by shocks}.  The H$\alpha$ velocity field shows evidence of a rotating disk, which
has a velocity amplitude of $\sim$240~km\,s$^{-1}$ at $\sim$ 136~pc from the nucleus. Assuming a Keplerian approach we estimated that the mass inside this radius  is $1.9\times10^9$ $M_{\odot}$, which is in agreement with the value obtained through the M-$\sigma$ relation,  $ M_{SMBH}=1.8\pm1.6\times10^{9}\,M_{\odot}$. Modelling the ionized gas velocity field by a rotating disk component plus inflows towards the nucleus along  filamentary structures, we obtain a mass inflow rate of $\sim$0.4~M$_\odot$. This inflow rate is enough to power the central AGN in NGC~5044.}

\end{abstract}

\begin{keywords}
  galaxies: individual (NGC\,5044) --
  galaxies: LINERs -- galaxies: stellar content -- galaxies: kinematics and dynamics --
  
\end{keywords}

\section{INTRODUCTION}

It is known for a long time that many elliptical galaxies show disks, as well as filaments of ionized gas 
{ \citep{macchetto96, ferrari99, tran01}.  Many of the early type emission line galaxies show characteristic spectra of a low-ionization nuclear emission-line region \citep[LINER,][]{heckman80}. 
However, the ionization mechanisms of this gas is still matter of debate. For example, very recently, \citet{ricci15} studied early-type galaxies with ionized gas using optical Integral Field Spectroscopy (IFS) and found that the  H$\alpha$ flux along the disk cannot  be  explained only  by ionizing photons emitted by an Active Galactic Nuclei (AGN). Other mechanisms, like ionization by shock \citep{heckman80, dopita95, allen08, dopita15} or Post Asymptotic Giant Branch (Post-AGB) stars \citep{stasinska08, singh13}, can also explain the observed spectra. This may suggest that the gas is being ionized by more than one ionization mechanism. }

Furthermore, large amount of molecular gas  (up to 10$^{7}\rm M\odot$ revelled by CO(2-1) 1.3 mm and H$_2$
2.121\,$\mu$m  emission) was found for a sample of bulge-dominated galaxies with dust lanes, revelling the complexity of these galaxies \citep{david14,davis15,strong88}. It was suggested that this molecular and atomic gas would be accreted through a recent merger with a gas rich galaxy and/or that it originates from a cooling flow \citep{sullivan15,david14}. { In addition, interstellar dust, mixed with the molecular and atomic gas is observed in early type galaxies \citep[e.g.][and references therein]{davis15}. Disks and dust filaments at kpc scales (with mass between 10 and 200 ${M}_{\odot}$ and Mid-Infrared luminosity up to 14.6 $\times$ 10$^8$ ${L}_{\odot}$) following the ionized gas distribution were also reported in early type galaxies \citep{ferrari99, ferrari02, goudfrooij94a}.}

Nowadays, galaxies with LINER-like spectrum are defined based on the optical line ratios { \citep{ho08}}: [OI]$\lambda$6300 / [OIII]$\lambda$5007 $\geq$ 1/3, [NII]$\lambda$6583 / H$\alpha$ $\geq$ 0.6, [OIII]$\lambda$5007 / H$\beta$ $\leq$ 3. In addition, over the last 30 years the most accepted explanation was that these objects are powered by weak AGNs
\citep{ferland83,halpern83}. However, in the last few years contradictory results were found between the
AGN-ionization scenario and predicted emission line strengths \citep{cid11}
or { with the spatial distribution of the ionized gas regions} \citep[e.g.][]{sarzi10, yan12, martins13,belfiore15}. In addition, using the Calar Alto Legacy Integral Field Area \citep[CALIFA,][]{sanchez12} survey of spatially resolved data \citet{singh13} have shown that galaxies with LINER-like spectrum are powered by hot old stars and accreting black holes are rare.  { In this work we propose to test this hypothesis by mapping in details the emission gas and stellar populations in the inner kiloparsec in a key object -- the gas rich early type galaxy NGC\,5044.}

NGC~5044 was observed over almost the entire electromagnetic spectrum, presenting rich structures, such as gas and dust filaments { \citep{ferrari99}}. It presents a very bright ionized gas emission in form of filaments extending up to 10 kpc with a complex kinematics { \citep{macchetto96}}. { Its radial velocity is blueshifted with respect to the systemic stellar velocity and the inner parts of the stellar velocity curves presents counter-rotation with respect to the galaxy outer regions  \citep{caon00}}.

{ The gas ionization mechanism of NGC5044 has been the aim of several studies,} for example,  \citet{rickes04} suggest that the gas in the central region ($\sim$4$\arcsec$) of NGC~5044 is ionized by a Low Luminosity AGN (LLAGN), while the outer regions are explained by post-AGB stars \citep[see also][]{macchetto96}. In addition, the overall stellar emission of this galaxy seems to be dominated by old stars over a large wavelength range, from UV \citep{annibali07, marino11}, optical \citep{rickes04} and Mid-infrared observations \citep{Vega10}. Unusual polycyclic aromatic hydrocarbons (PAH) emission was detected in Spitzer-IRS observations by \citet{Vega10}, this emission was not expected in this galaxy because PAH is a typical feature in star-forming galaxies. Moreover, the IRS spectrum of this object displays strong H$_2$ molecular and atomic gas emission \citep{Vega10,panuzzo11}. { Furthermore, this galaxy shows several X-ray cavities within the inner 10~kpc, which are attributed to multiple AGN
outbursts over past 10$^8$ yrs \citep{david09,gastaldello09}}. Moreover, \citet{david14} using Atacama Large
Millimeter/submillimeter Array (ALMA) data argue that the detection of 230 GHz continuum emission shows that it is currently going through another outburst, probably due a recent accretion event. In addition, these authors report the detection of cold molecular CO gas moving towards the center of this galaxy, and suggested that it can be interpreted as the presence of a cooling flow in the galaxy center direction.

{ We present in this paper for the first time the optical IFS obtained with  Gemini Multi-Object Spectrograph (GMOS) in integral field unit (IFU) of the central 1 kpc of NGC~5044, whose gas ionization source is still controversial. Beside that, there is no information about its central gas kinematics.
Although strong evidence suggests possible presence of a contribution of an AGN, other energy source also seem to be present in this galaxy ionizing their extended filaments \citep{macchetto96,rickes04}. This data allowed us to spatially resolve the inner kpc of this source and to map the possible ionization mechanisms through the analysis of stellar populations and emission line flux ratios. The goals} of this work are to establish the star formation history (SFH), the nature of the energy sources that ionizes the gas and search for the existence of inflowing or outflowing gas associated with the AGN and/or other mechanism, like cooling flows, for the central kpc of NGC\,5044.

This paper is structured as follows:
In Sec. 2 we described the observations and data reduction procedures. Details of the spectral synthesis method and its results are presented in Sec. 3. In Sec. 4 we analyse the gas kinematics. The emission gas and its ionization mechanism is studied in Sec. 5.  We show the velocity field model and estimates the black hole mass in Sec. 6. The geometry and the origin of the observed gas is discussed in Sec. 7. Lastly, we present our conclusions in Sec. 8.

\section{OBSERVATIONS AND DATA REDUCTIONS}

{ NGC~5044 was observed on May 04, 2013 with the Gemini South Telescope using the  GMOS in the IFU mode \citep{allington02, hook04} using the one slit setup.} In this configuration the spectrograph uses 500 hexagonal micro-lenses centered on the object, and other 250 micro-lenses, separated by 1 $arcmin$ from the object, to simultaneously measure the sky emission. These micro-lenses, located at the focal plane of the telescope, divide the image in slices with a diameter of 0.2$\arcsec$ and are connected to a set of fibers. This setup produces a data cube with one spectral dimension and two spatial dimensions with a field-of-view of 3.5 x 5 $arcsec^2$. We used the B600-G5323 grating, with a central wavelength of 5620$\AA$. The spectra covered a
range from 4260 to 6795 $\AA$ with a 1.7$\AA$ resolution, obtained from the full width at half maximum (FWHM) of CuAr arc lamp lines.  In Table \ref{table:prop} we list some of the general parameters of NGC\,5044.
 Flat-Field exposures, bias and CuAr lamp spectra were obtained for calibration and correction of the data cubes. 
{ The DA white dwarf EG~274 \citep{hamuy92} was used to perform  the spectrophotometric calibration.} 
The seeing value of the observation was measured in the acquisition r image obtained with the GMOS imager (SDSS system). This value and other observation parameters are shown in Table \ref{table:obs}.

\begin{table} 
\caption[Table caption text]{Basic parameters of NGC~5044.}
\begin{center}
\begin{tabular}{|l|l|} \hline \hline
Parameter & NGC~5044 \\ \hline
RA (2000) & 13h 15m 24.0s   \\  
DEC &  -16$^{\circ}$ 23' 09"  \\ 
Morphology & E0  \\ 
B (mag) & 11.83   \\ 
M$_{B}$ (mag) & -22,39  \\ 
A$_{v}$ (mag) & 0.192  \\ 
Radial velocity$ ^{a}$ (km s$^{-1}$) & 2782  \\ 
Distance (Mpc) & 37  \\  
{ Redshift (z)} & 0.009280 \\
{ L$_{x}$ $^{b}$ (erg s$^{-1}$)} & 6.8e+39  \\ \hline
{ Data available in NED\tablefootnote{The NASA/IPAC Extragalactic Database (NED) is operated by the Jet Propulsion Laboratory, California Institute of Technology, under contract with the National Aeronautics and Space Administration.}}\\
\multicolumn{2}{l}{ $^{a}$ Radial velocity measured by \citet{ogando08}.} \\
\multicolumn{2}{l}{ $^{b}$ X-ray luminosity in 0.3–8 keV \citep{liu11}.}  \\ \hline
\end{tabular}
\end{center}
\label{table:prop}
\end{table}

\begin{table} 
\caption[Table caption text]{Observation log.}
\begin{center}
\begin{tabular}{|l|l|} \hline \hline
Parameter & NGC~5044 \\ \hline
Observation date & May 04 2013   \\  
Gemini Programme &  GS-2013A-Q-52  \\ 
Seeing (arcsec) & $\sim$ 0.7  \\ 
Airmass & 1.14  \\ 
T$_{exp}$(s) & 1800  \\  \hline
\end{tabular}
\end{center}
\label{table:obs}
\end{table}

{ We followed standard procedures to reduce the data, using {\sc iraf} \citep{tody86,tody93} and the tasks contained in the {\sc gemini iraf} package, as described in detail in \citet{ricci14}.}
The data cube was built with a spatial sampling of 0.05$\arcsec$, resulting in a FoV with 66$\times$98 spaxels. The high spatial frequency noise was removed with a {\sc butterworth filter} { \citep{Gonzalez, ricci14} using order n=6 }and a cut-off frequency Fc=0.13 FNy, where FNy = 0.5 spaxel$^{-1}$ is the Nyquist frequency. This cut-off frequency was chosen { to remove only spatial frequencies higher than the seeing.} 
 { In addition, the Principal Component Analysis (PCA) tomography \citep{heyer97, steiner09} was applied in order to remove signatures of low spatial and spectral frequency \citep{ricci14}.} 
{ To correct the atmospheric refraction effects, we applied the equations proposed by \citet{bonsch98} and \citet{filippenko82}.}

The data cube spectra was corrected for Galactic extinction using the \citet{cardelli89} extinction curve (CCM) for an $A_{v}=0.192$ \citep{schlafly11} and  by Doppler effect using the radial velocity of 2782 \kms given by \citet{ogando08}. { These authors derived the radial velocity through the cross-correlating
technique by using the RVSAO package \citep{kurtz98} with stellar spectra of G and K giant stars.}

\begin{figure}
\centering \includegraphics[width=.5\textwidth]{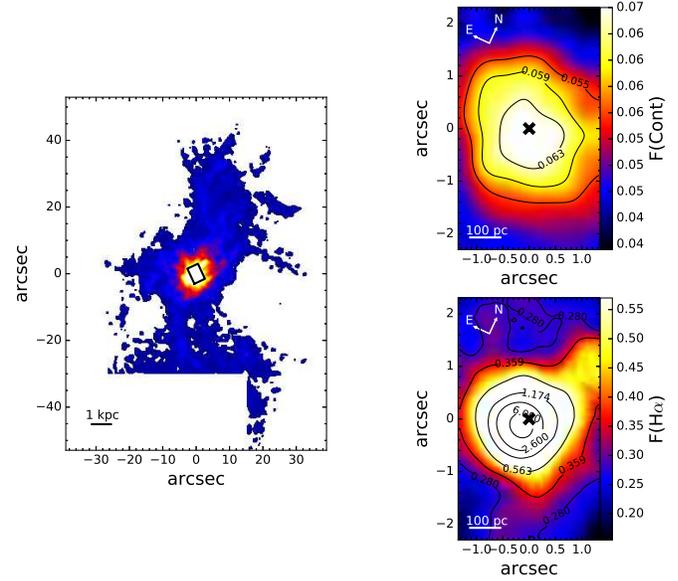} 
\caption{ {Flux maps for NGC~5044. Left: H$\alpha$+[NII] emission obtained with NTT+EFOSC2 \citep{macchetto96}. The black rectangle represents the GMOS-IFU FoV (5 x 3.5 $arcsec^2$). Upper right: GMOS-IFU mean continuum image obtained in the spectral range: 4730-4780$\AA$. Lower right: GMOS-IFU H$\alpha$ map. GMOS maps are in units of $10^{-15}$ erg $cm^{-2}$ $s^{-1}\AA^{-1}$.}}
\label{gal} 
\end{figure}

We show in the left panel of the Figure~\ref{gal} a H$\alpha$+[NII] emission image of NGC~5044 { obtained with NTT+EFOSC2} \citep{macchetto96}, where the black rectangle in the middle represents the GMOS FoV. In the right panel we presented two images extracted from our data cube. The continuum image was obtained by estimating the mean value over the wavelength range between 4730 and 4780 $\AA$, with the continuum peak defined as the photometric center of the galaxy. We also show the H$\alpha$ image with contours drawn to highlight the extended ionized gas filaments. It is clear from these maps that the morphology of the continuum and H$\alpha$ are distinct and suggests that the ionized gas follows the larger kpc scale filamentary structure, while the continuum follows the stellar distribution.

{ The wavelength calibration uncertainty was estimated from the arc lamp lines (see section~\ref{kin}). The spectroscopic standard star was not observed at the same night as the galaxy and this may be affected the accuracy on the absolute flux determination, given 30\% of uncertainty in their values. It is worth mentioning that the emission line flux ratios, used to build the diagnostic diagrams were not affected by the systematic flux uncertainty, since each used ratio were measured over nearby emission lines. Nevertheless, we check uncertainties for [NII]$\lambda$6584/H$\alpha\lambda$6563 and [OIII]$\lambda$5007/H$\beta\lambda$4861 ratios in several spaxels and found a deviation of 3\%. The instrumental dispersion was obtained from the FWHM of the CuAr arc lamp, { and we used it to corrected the velocity dispersion}. We noted that the deviation are larger in blueward than in redward, given an uncertainties of $\sim$8\% and $\sim$4\% respectively.}

\section{Stellar Population Distribution} \label{synt}

In order to obtain the SFH and to analyse the gas emission free from the stellar population contamination, we performed stellar population synthesis spaxel by spaxel. Therefore, we use the \str\ code \citep{cid04,cid05,mateus06,asari07,cid08}, which essentially fits the whole underlying observed spectrum, $O_{\lambda}$, with a model spectrum, $M_{\lambda}$. 
To create a model spectrum, $M_{\lambda}$, the code finds a combination of $N_{\star}$ simple stellar populations (SSPs) resulting in a final population vector { x}, whose components represent the fractional contribution of each SSP to the total synthetic flux at wavelength  $\lambda_0$ \citep{cid04,cid05}. Extinction is parametrised by the $V$-band extinction $A_V$ and modelled by \str\ as due to foreground dust. In summary the code solves the following equation: \\

\begin{equation}
{\it M_{\lambda}= M_{\lambda 0} \left[ \sum_{j=1}^{N_{\star}} x_j b_{j, \lambda} r_{\lambda} \right] \otimes G(v_{\star}, {\sigma}_{\star})},
\end{equation}
where $M_{\lambda 0}$ is the synthetic flux at the normalisation wavelength, $G(v_{\star}, {\sigma}_{\star})$ is a Gaussian function used to model the line-of-sight stellar velocity distribution, which is centered at velocity $v_{\star}$ with dispersion $\sigma_{\star}$. The term $x_j$ is the {\it j}th population vector component of the base set; $b_{j, \lambda}$. The spectra of the SSP of the base of elements are normalized at $\lambda_0$, the reddening term is represented by $r_{\lambda}=10^{-0.4(A_{\lambda}-A_{\lambda 0})}$. The final fit is carried out by minimizing an $\chi^2$ equation. The robustness of the fits can be measured by the \str\ output parameter ADEV, which is the average deviation over the fit of pixels $|O_{\lambda}-M_{\lambda}|/O_{\lambda}$.

It is clear from the above that the most important ingredient in stellar population synthesis are the SSPs used in the fit (what we call  base set). We used SSPs taken from \citet{bc03} models. These models were chosen because they are the most adequate to fit our data, since they have a spectral resolution of 3.0 \AA, that is close to our data spectral resolution (see section~\ref{kin})  and do offer enough spectral coverage (3200 - 9500\AA) to fit our data. Beside they do provide an adequate number of SSP in order to have all the possibilities of ages and metallicities found in galaxies. { In addition, we decided to use the BC03 models because they are widely used, thus allowing for comparison with published results. These models are computed using the isochrone synthesis code developed by BC03 and are divided in 221 spectra 
with ages between 0.1\,Myr - 20\,Gyr for a wide range of metallicities. These models mainly use the BaSeL 3.1/STELIB libraries, with a spectral resolution of 3\AA\ (STELIB) in the optical region.}  Following \citet{dametto}, our final base set was selected in order to have enough, representative and not degenerated elements, covering 7 ages ($t=$ 0.6, 0.9, 1.4, 2.5, 5, 11 and 13 Gyr) and three metallicities (0.2~\,Z$\odot$, 1~\,Z$\odot$, 2.5~\,Z$\odot$). In the fits { the spectral resolution of the data was resampled to match the models.   The} \citet{cardelli89} extinction law, which is typically used to fit the extinction data both for diffuse and dense interstellar medium. Since we are fitting a LINER-like spectrum galaxy, we additionally included a featureless continuum (FC) in the form of a power law ($F_\nu\propto\nu^{-1.5}$) in order to account for a possible non thermal contribution \citep[e.g.][and references therein]{cid05,riffel09}.  

To map the whole FoV, the stellar population fits were performed for each one of the spaxels.  The nuclear region ($\sim$200 pc) was defined as the one where a broad component was detected on H$\alpha$ emission. Fig.~\ref{5044_spec} shows the fit of the spectrum at the nuclear spaxel, defined as the location of the continuum peak. The emission lines and the final continuum fit are also shown. The percentage contribution of each age (in mass\footnote{The mass fractions are obtained using the M/L ratios for each SSP, see {\sc starlight} manual for details} and light fractions) and FC to the final fit are presented in the bottom panels.  Here we discuss only the light fractions, since it is the parameter we derive directly and where larger differences are observed (in terms of mass, the emission is completely dominated by the old stellar population which has an high M/L).

The quality of the fit over the whole FoV can be seen in Fig.~\ref{adev} where the parameters ADEV and $\chi^2$, are shown. In this figure we do also show a signal to noise ratio (SNR) map. It is clear from this map that even for the borders of the FoV a SNR $>$15 is found. It is worth mentioning that according to tests presented by \citet{cid04}, for spectra with values of SNR $>$ 10 \str\ does properly reproduce the input parameters, thus producing reliable results for the stellar populations fits. {  However, the strengths of the Mg$\lambda$5143 and Na$\lambda$5894 lines are not well reproduced by our fitting, possible due to the fact that the BC03 models do not consider $\alpha$ enhancement processes. In addition, the NaD doublet has contribution of the ISM. Due to this fact we have tested the stellar population { fits in three cases, with and without removing both the NaD and Mg$\lambda$5143 lines form the fit and only removing the  NaD doublet}. We have not found any significant difference between both fits. In both cases, less than $\sim$ 10\% of these lines strength was not reproduced by the fitted model. Likely the NaD residual seen in Figure ~\ref{5044_spec}, is due to ISM contribution present in the nucleus of NGC~5044 \citep[see][for example]{ferrari99}. { As discussed in \citet{davis12} the blueshifted NaD doublet provides a good probe of the cold interstellar medium outflow. In the case of NGC~5044 we have not found any blue-shifted component for NaD doubled in all our FoV, indicating that in this galaxy, the cold material is probably not outflowing.}

\begin{figure}
\begin{minipage}[t]{0.5\textwidth}
\includegraphics[width=\textwidth]{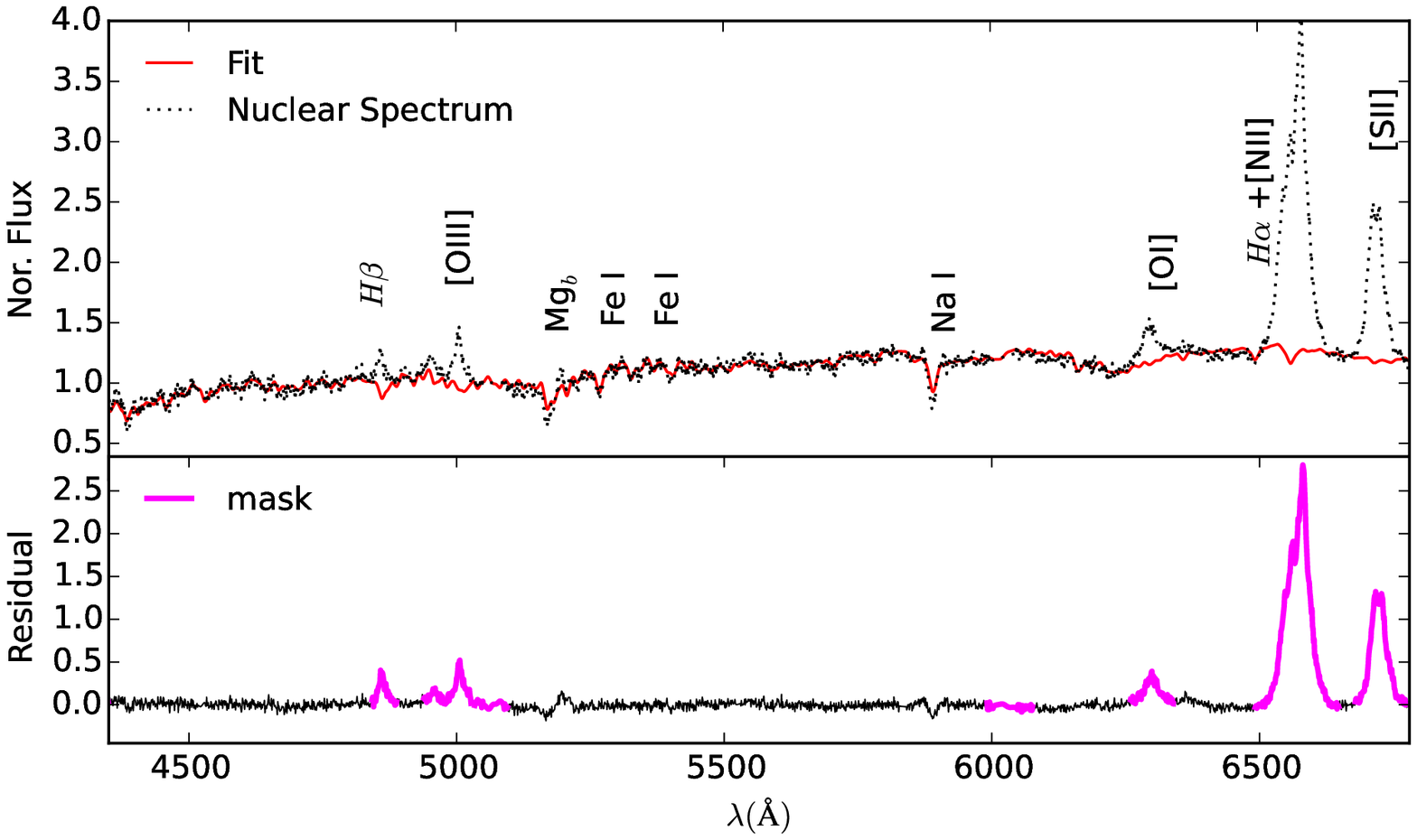}
\end{minipage}
\begin{minipage}[t]{0.5\textwidth}
\includegraphics[width=\textwidth]{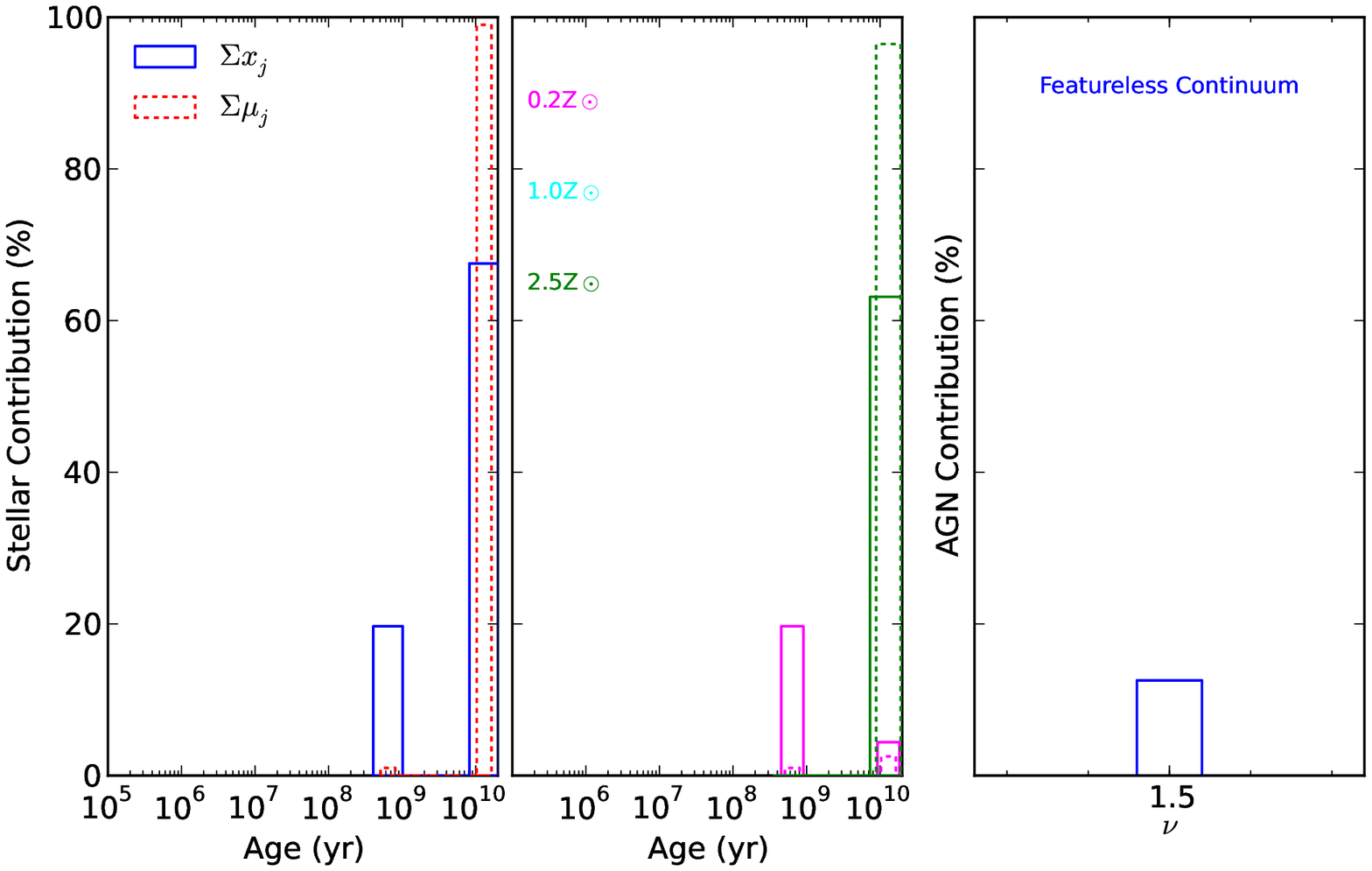}
\end{minipage}
\caption{Example of fits. Top panel: the observed and modelled spectra normalised to unity at 4755\AA, central panel: residual spectrum. The masked regions are in magenta and the residual spectrum in black lines.  At the bottom we show the contribution of the stellar population vectors in light fractions ($x_j$, full line) and mass fractions ($\mu_j$, dashed line). Left: mettalicities summed contributions, center: contributions separate by each metallicity, { right: featureless continuum contribution.}\\}
\label{5044_spec} 
\end{figure}

\begin{figure}
\centering \includegraphics[width=.5\textwidth]{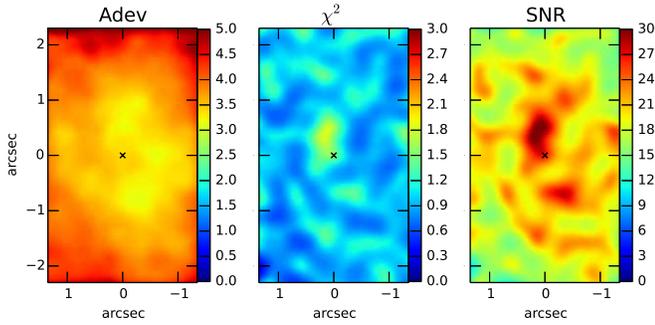} 
\caption{Quality of the fit maps. Left: percent mean deviation, 
center: $\chi^2$ and right: SNR. For details see text.\\}
\label{adev} 
\end{figure}

Interestingly, our fitting resulted only in contributions of four of the base components: three SSPs with ages 900\,Myr (0.2~$Z\odot$) and 13\,Gyrs (0.2 and 2.5~$Z\odot$), plus a FC component. The 2D maps for these components (in light fractions), together with mean age, A$_v$, FC and mean metallicity are shown in Fig.~\ref{pop}. It is clear from this figure that the light in the central region of NGC~5044 is dominated by an old stellar population. Besides, it is also clear that an intermediate age (900\,Myr) stellar population is present, surrounding the galaxy nucleus. { As far as we know, the presence of stellar populations with these ages (900~Myr) in the nuclear region of elliptical galaxies is not a common trend. For example, from an extensive stellar population study performed with large field data of the CALIFA survey, \citet{gonzalez15}, found that in average  the S0 and E  galaxies have a dominant stellar population with almost the same age and they are older than those found in spirals. In addition, they reported for elliptical galaxies a small variation on the stellar ages along the galaxy radii (from 10$^{10}$ to 10$^{9.7}$ yr from the center to 3 Half Light Radius, \citep[see figure 17 of][]{gonzalez15}. The differences in the results found by \citet{gonzalez15} and ours may be due to the spatial and spectral resolution probed by GEMINI GMOS. For instance, the presence of spatially resolved  circumnuclear intermediate age stellar populations, like we found in NGC5044, was reported for the bulges of spiral Seyfert galaxies \citep{riffel10, riffel11, storchi12}. The maps of mean age and mean metallicities shows a trend consistent with the stellar population becoming more metal rich and older with decreasing distance with respect to the galaxy centre.} Some studies have shown that the stellar population in early type galaxies also become progressively more metal rich and older with increasing velocity dispersion \citep{thomas05, clemens06, clemens09}. Indeed, we found higher $\sigma$ values in the central region of the NGC\,5044 (see Sec. ~\ref{kin}), thus our result does further suport the findings by these authors. 

In addition, our stellar population results are in agreement with the results found by \citet{rickes04}, who performed stellar population synthesis using optical long slit data in a much larger spatial scale than ours. These authors used two different bases of elements, one with three components with metallicities [Z/$Z_{\odot}$] = 0, -0.4 and -1.1 and age = $10^{10}$ years, and another with solar metallicity and ages with $10^{10}$yrs, $10^8$yrs and 10$^6$yrs. They noted that there was an improvement on the fit when considering different metallicities over the different ages, with the more metallic component dominating in the central region, while the lower metallicities were enhanced in the outer regions.

Literature results on the stellar content of NGC\,5044, as those presented by \citet{werner14} based on the GALEX and WISE data report an unusual star formation rate in NGC 5044 (SFR= 0.073 $M_{\odot}/yr$, using a region of 10\,kpc, over the last 1\,Gyr). Besides,  \citet{david14} suggested that the unusual lines of PAHs found in NGC 5044 and the uncertainty in the stellar age are consequences of episodes of star formation that occurred in the last 1\,Gyr \citep{marino11}. These results are further supported by our finding, that there is a significant contribution of the 900 Myr stellar population, distributed along the center of the galaxy (reaching values of $\sim$ 50\% at some locations).

\begin{figure*}
\centering \includegraphics[width=.9\textwidth]{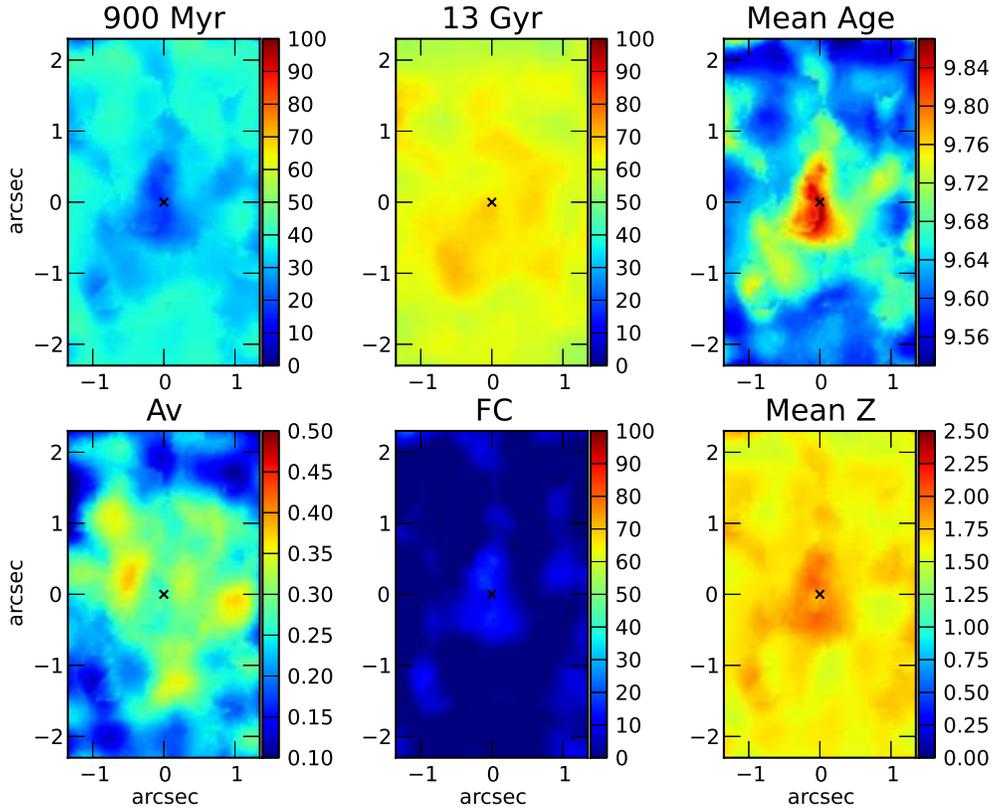}
\caption{Top panel shows the population maps of NGC~5044 at age, wherein in the left we can see the population contribution of 900~Myr, in the center is the population contribution of 13 Gyr and at right is the mean age. The bottom panel shows Av map at left, the FC contribution in the center and the mean metallicity (Z) at right}
\label{pop} 
\end{figure*}

We also found a FC contribution of about 20\% in the nuclear region. This component is consistent with a scenario where the gas in the nuclear region is ionized by a weak AGN. However, we find some significant contribution of the FC in the external regions. Since this component is degenerated with a young reddened stellar population we interpret this contribution as being related with the fact that BC03 models do not include blue horizontal branch stars, and {\sc starlight} fits tend to choose a young population to account for this blue population \citep[see][for details]{cid10b}.

{ The visual extinction map (Av) shows a clear enhancement at the galactic centre, reaching $\sim$0.35\,mag and decreasing to 0.1 mag at the external region. \citet{ferrari99} found a mean extinction Av=0.01 (mag) within a region of $10^{\prime\prime}$, which is about twice of the size of our region. From this Av, they estimated a dust mass of about 10$^{4}$M$_{\odot}$ distributed irregularly along the galaxy. In addition, a hot dust mass of 67 M$_{\odot}$ was determined by \citet{ferrari02} from Mid-IR data. }

%\section{kinematics of ionized gas: radial velocity and velocity dispersion}\label{kin}
\section{kinematics of ionized gas}\label{kin}
\label{sec_kine}

In order to map the gas kinematics we measured the centroid velocity (hereafter velocity field, ${v_{r}}$) and velocity dispersion ($\sigma_{v_{r}}$) from the data cube without the subtraction of the stellar populations contribution { (see Sec.~\ref{synt})}, because the stellar population templates have slightly lower resolution than the data cube, this difference would affect the accuracy on velocities determination. 
We have used an in house fitting code (Hernandez-Jimenez et al. - in preparation) to fit Gaussian functions to the emission-line profiles. We fitted single- or multi-Gaussian components to the spectra in order to obtain the flux of each emission line. The line pseudo-continuum is linearly fitted using side-bands regions, free from emission or absorption features, as close as possible to the emission line. 
The uncertainty associated to the measurements of the velocity field ($\delta_{v_{r}}$), in a given line at $\lambda_0$, depends implicitly 
on the error of the Gaussian peak centroid,  ($\delta_{\lambda_{0}}$). Then, we do a straightforward error propagation of the basic expression for the Gaussian peak centroid ($\lambda_0$) to calculate $\delta_{v_{r}}$. We derived  the following expression:

\begin{equation}
\centering                                                                
\delta_{v_{r}}=\frac{c}{\lambda_0}\delta_{\lambda_0};\,\,\delta_{\lambda_0}= \frac{I_{0}\sqrt{N}}{F_{\lambda}} \delta_{\lambda},
\end{equation}  
where $I_{0}$  is the intensity at maximum of the fitted Gaussian, $N$ is the number 
of pixels, $F_{\lambda}$ is the line flux, and  $\delta_{\lambda}$ is the error of
the spectral dispersion taken from the FWHM (1.7\AA) of the CuAr arc lamp. The error map $\delta_{v_{r}}$ of 
H$\alpha$ line is shown in panel $(d)$ of Fig. \ref{model}, where we can see that the uncertainties in  velocity field in the nuclear region is lower than 10 \kms, while at the data cube border (where the SNR is lower), the error is $\sim$ 20~\kms.

\begin{figure*}
\centering \includegraphics[width=0.8\textwidth]{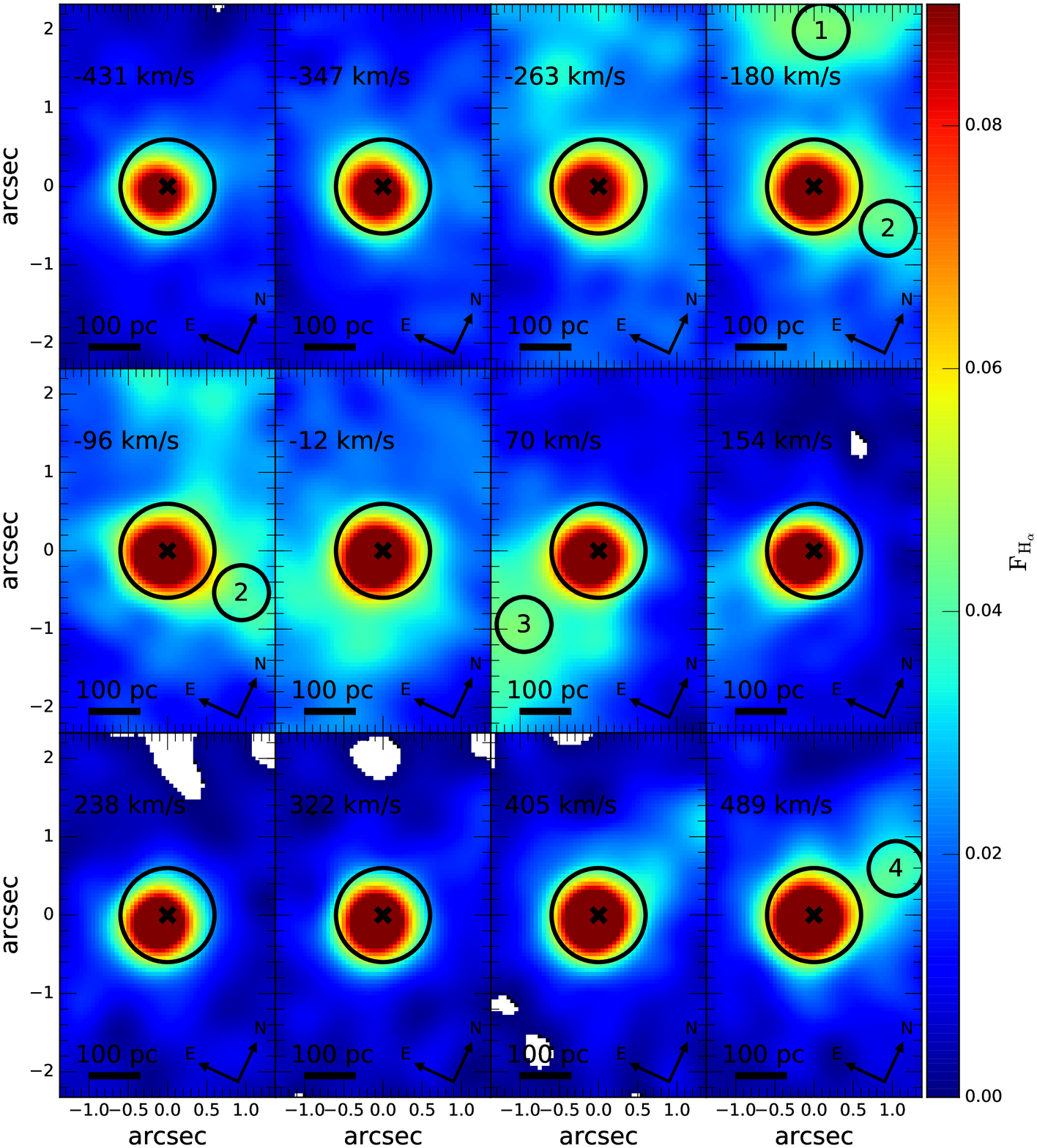}
\caption{Velocity channels along the H$\alpha$ emission line in $\sim$ 80 
             \kms\ bins centered on the velocities indicated in each panel. 
              The intensities are represented in the color scale to the right 
              in flux units of $10^{-15}$ erg $\rm cm^{-2}$ $s^{-1}$.  
              { The central circle in all channels represents the nuclear region.
              The photometric center of the galaxy is marked with a symbol ``x'' in all channels.}}
\label{Channelmaps} 
\end{figure*}

In order to study the kinematic components of the ionized gas in the circumnuclear
region of NGC~5044, we obtained channel maps between { -431\,\kms\ and +489\,\kms\,  relative to the systemic velocity of the galaxy derived by modelling the gas velocity field (see Sec.~\ref{IonG} and Table~\ref{table:disk}), in velocity bins of 80\,\kms\, (see Figure~\ref{Channelmaps}). We identified, in the negative 
channel map -180 \kms, a bright structure at NE, named as region 1, and a weak structure (region 2) at west, which is also observed at the channel -96 \kms. In the positive channel maps the most conspicuous structure was named as region 3 and is located at South, at the channel 70 \kms. There is also a filamentary structure, region 4, located at North 
between 405 and 489 km s$^{-1}$.} Comparing these ionized gas structures with the ones of molecular gas 
identified by \citet{david14}, using ALMA observations, we found that the brighter
peak located at NE of the channel map -180 km s$^{-1}$ match with the
approaching cloud 13 \citep[see Figure 6 in][]{david14}, whereas the south filament in 
channel map 70 km s$^{-1}$ coincides with the receding molecular gas 
cloud 18. In both regions \citet{david14} found evidence that the molecular gas is 
falling into the center of the galaxy. On the other hand, we point out that 
the central region appears in all channel maps, indicating the presence of a nuclear broad component with a FWHM $\sim$ 3000 km s$^{-1}$, typical of AGN.

\begin{figure*}
\begin{minipage}[t]{0.49\textwidth}
\includegraphics[width=\textwidth]{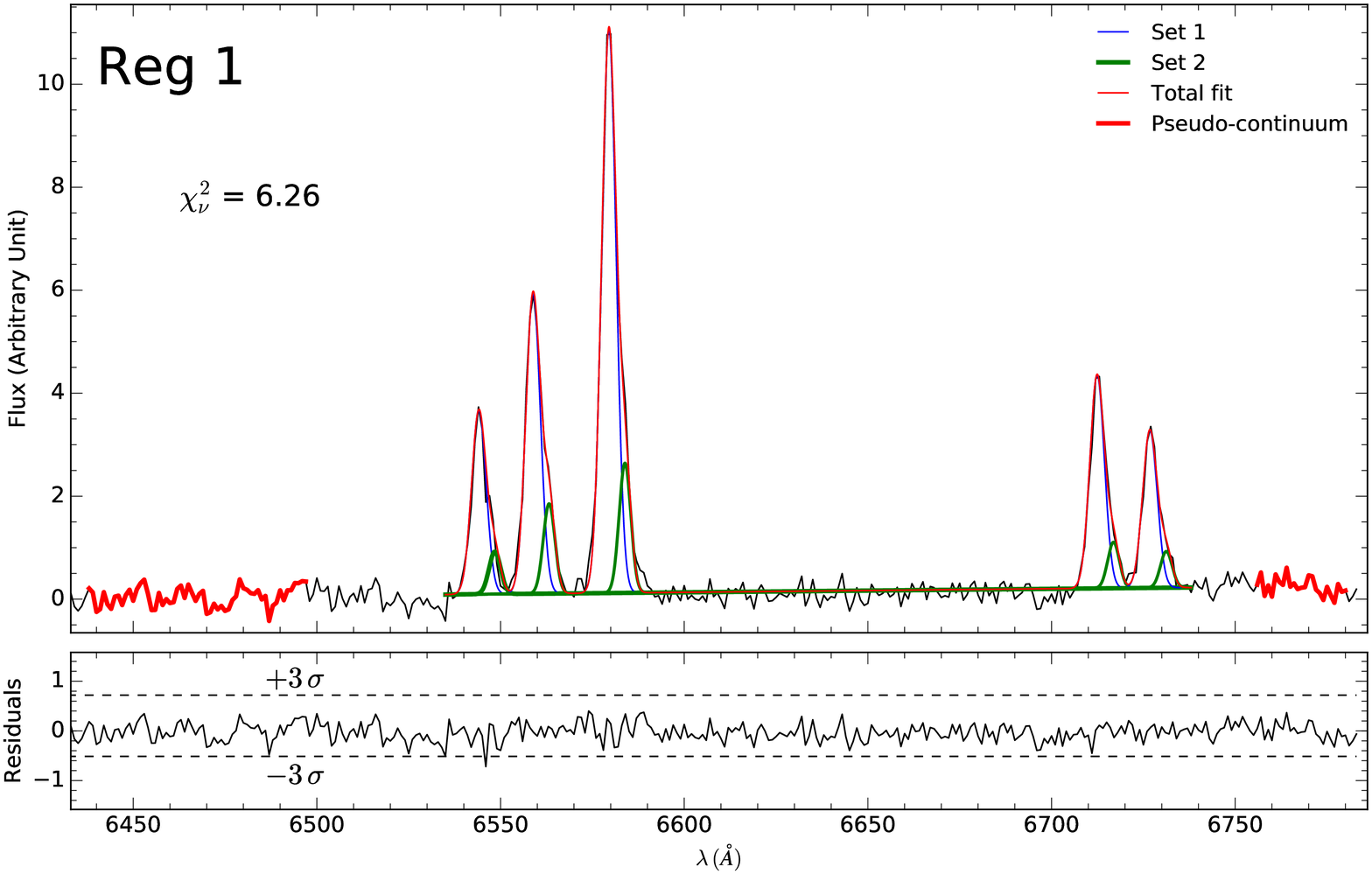}
\end{minipage}
\begin{minipage}[t]{0.49\textwidth}
\includegraphics[width=\textwidth]{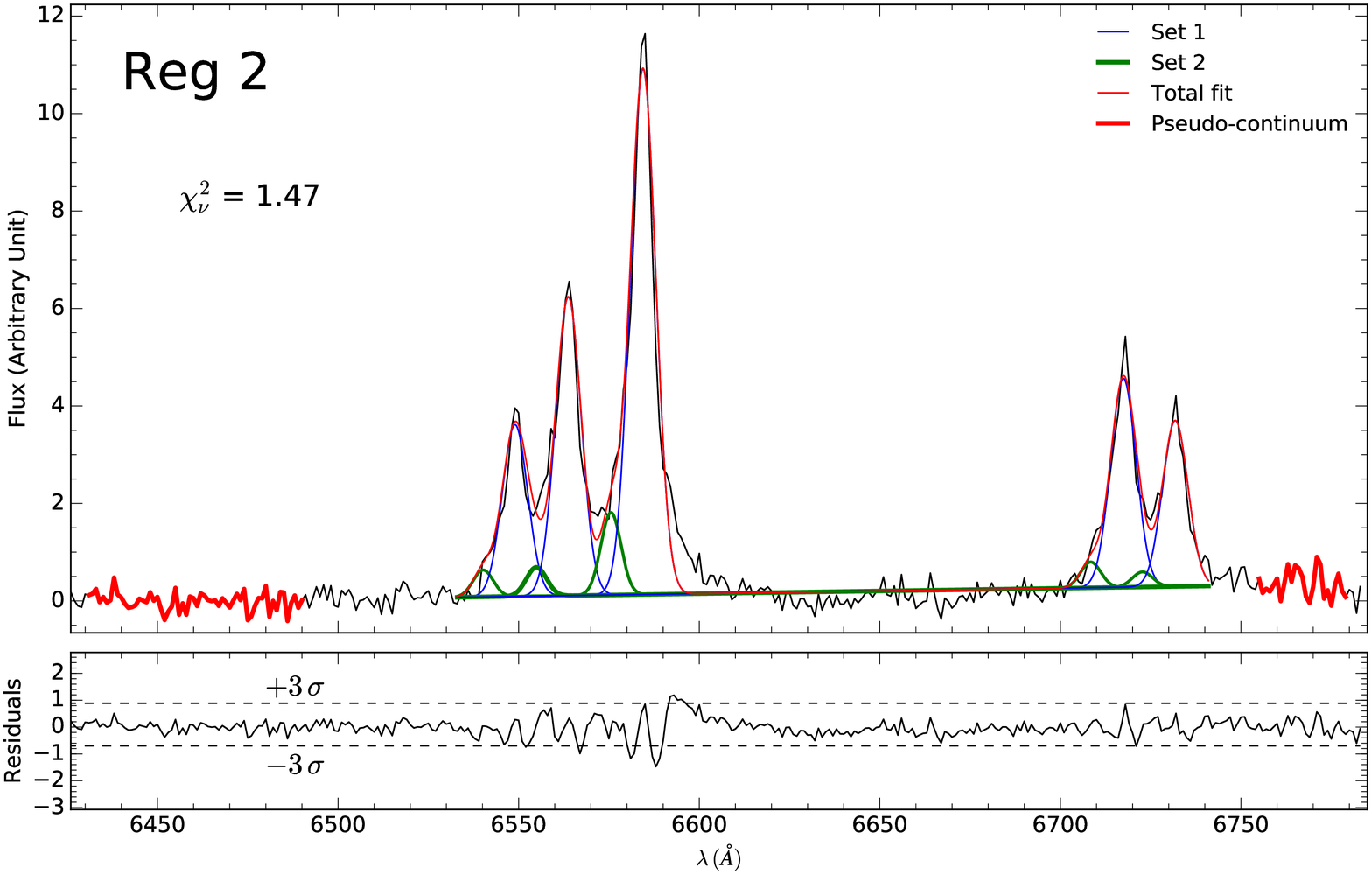}
\end{minipage}
\begin{minipage}[t]{0.49\textwidth}
\includegraphics[width=\textwidth]{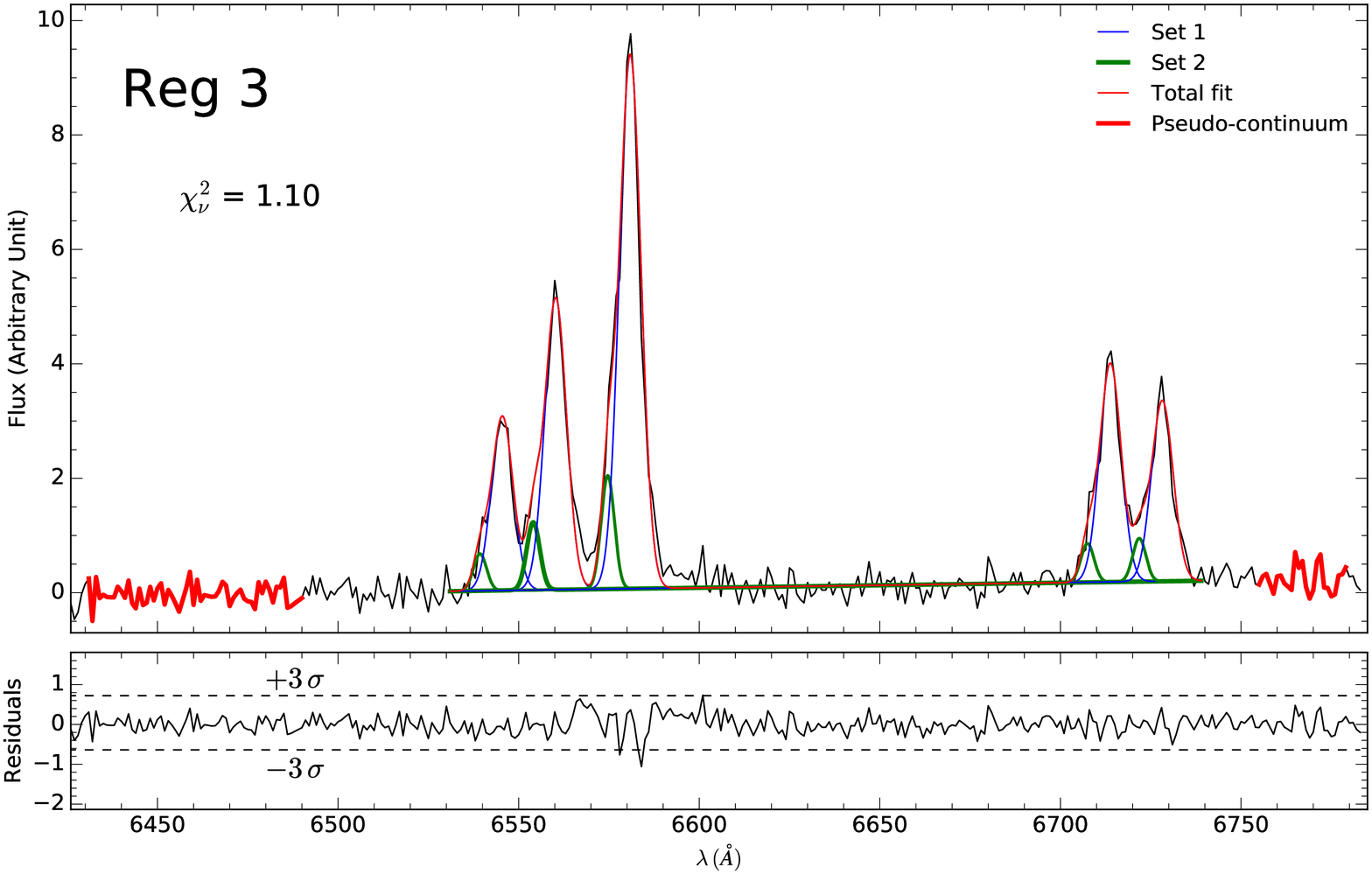}
\end{minipage}
\begin{minipage}[t]{0.49\textwidth}
\includegraphics[width=\textwidth]{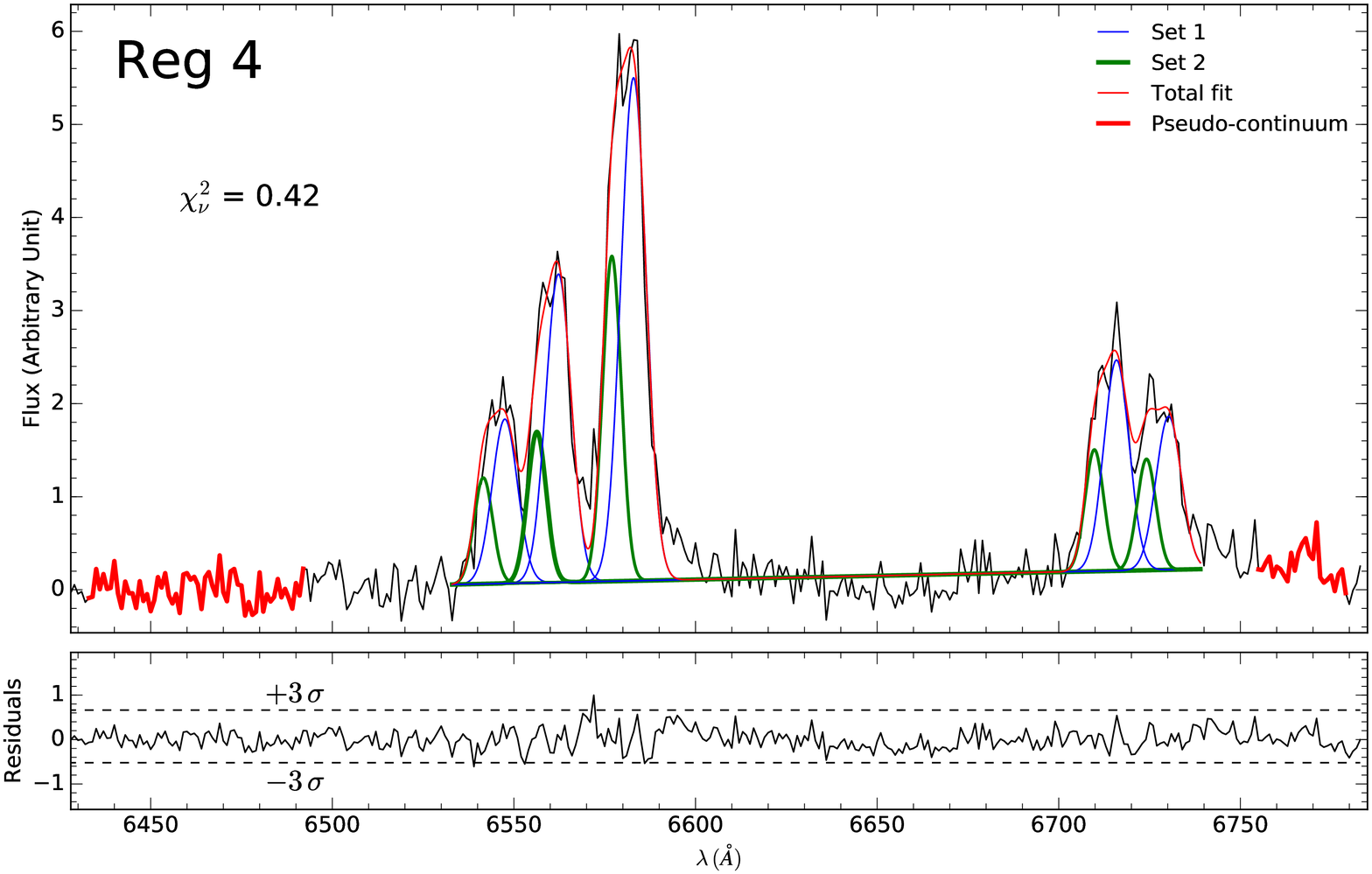}
\end{minipage}
\begin{minipage}[t]{0.49\textwidth}
\includegraphics[width=\textwidth]{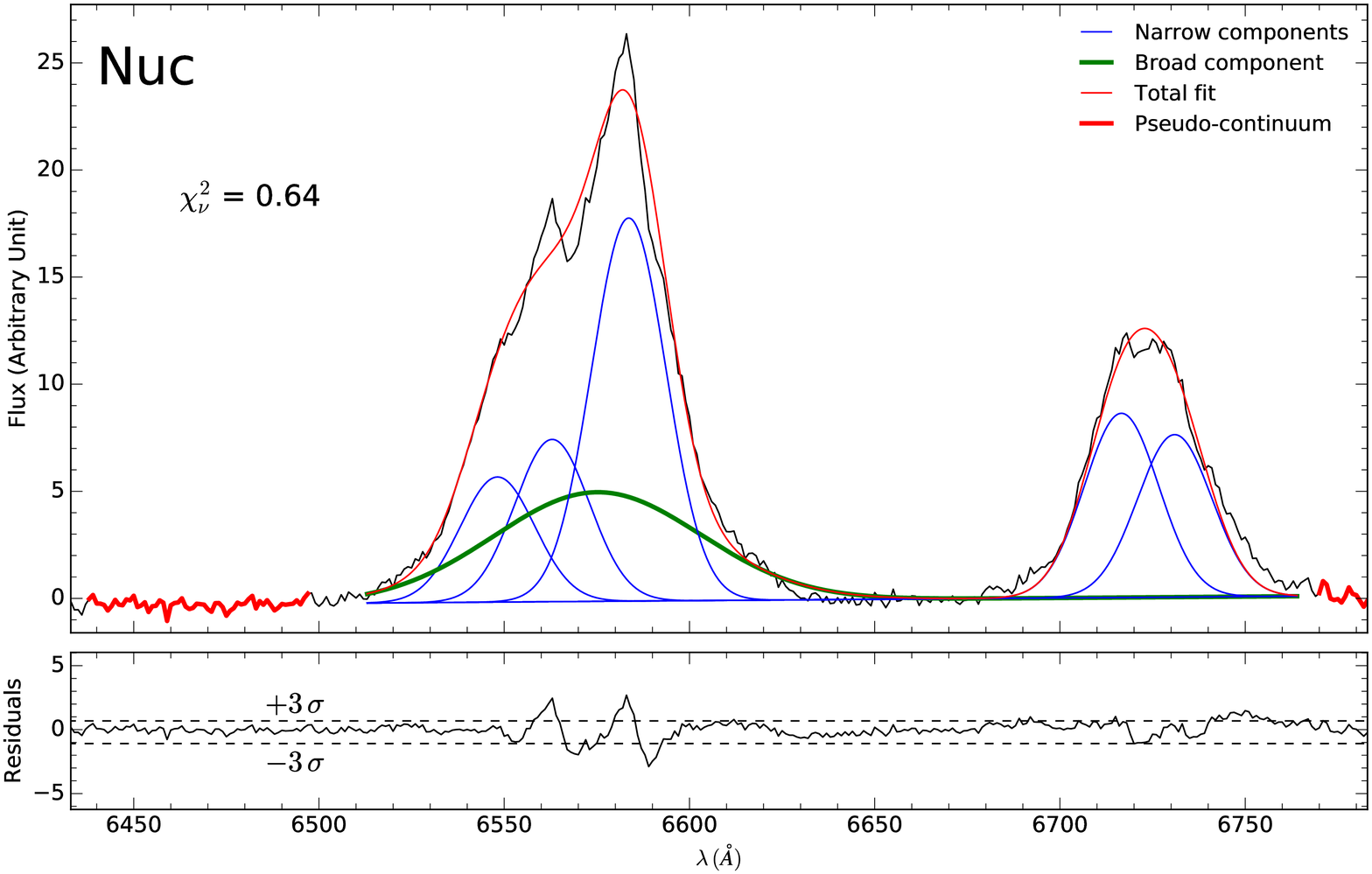}
\end{minipage}
\caption{Line profiles of [NII]$\lambda$6548, H$\alpha$, [NII]$\lambda$6583, [SII]$\lambda\lambda$6717,31  obtained in the apertures of $\sim$0.7$\arcsec$. The profile of the nuclear aperture shows components with FWHM $\sim$1080 \kms and a H$\alpha$ broad component (FWHM $\sim$3000 \kms). The emission line profiles obtained in the other regions are composed by two set of narrow lines (FWHM $<$400 \kms).}
\label{espectros} 
\end{figure*}

{ We fitted the [NII]+H$\alpha$ complex at the nuclear region with six gaussians, five corresponding to the narrow components of the lines [NII]$\lambda$6548, H$\alpha\lambda$6563, [NII]$\lambda$6583, [SII]$\lambda\lambda$6717,31 and one to account for the H$\alpha$ broad component.  We constrained the narrow components as having the same redshift and FWHM for all lines, and fixed [NII]$\lambda$6583/[NII]$\lambda$6548 = 3.1 \citep{osterbrock85}. The fit for the additional component was done with all parameters free, resulting in a large FWHM value ($\sim$3000~\kms), thus we associated it with a broad H$\alpha$ emission. To properly  reproduce the emission line  profiles of the outer regions (1, 2, 3 and 4), it was necessary two sets of gaussians, constrained as mentioned above, but each set with their own values for redshift and FWHM. We only considered lines with SNR above than 3$\sigma$ of the fitted continuum (see Fig.~\ref{espectros}). In Table~\ref{table:kin} we show the values obtained for  each fit, where ${V_{c}}$ is the peak velocity of the line, and $|\Delta V|$ is the shift between the line sets. The fitted H$\alpha$+[NII] and the [SII] lines profiles, obtained with apertures of the 0.7\arcsec, are shown in Figure~\ref{espectros}. The secondary set of lines of regions 2, 3 and 4 are blueshifted with respect to the first one. The difference in velocity between the sets for region 2 is 407\,\kms (the largest one). We point out that this region is the  closest from the nucleus. The velocity difference for regions 3 and 4, are $\sim$300 \kms. The secondary  set of lines for region 1 is redshifted with respect to the primary one in about 200\,\kms. Figure~\ref{espectros} also shows the line profiles for the nuclear region, where the H$\alpha$ broad component is blueshifted with respect to the narrow component by 558 \kms. The FWHM of these components are $\sim$1080\kms and $\sim$3000 \kms, respectively. { It is clear from Figure~\ref{espectros}, that the regions 2 and 4 do present a wider FWHM than regions 1 and 3. Besides they do show a weak red wing in the [NII]+H$\alpha$ complex (very close to the 3$\sigma$ detection limit). We have not a very clear explanation for this component, however, we do speculate that this weak component is most likely originated from material outflowing from the AGN, once these two locations are closer to the galaxy center.} Similar results  were found for other early type galaxies, for example, \citet{ricci14} fitted two sets of Gaussians to the emission lines in the inner 100~pc of 10 early type galaxies and found that for seven galaxies one of these sets has FWHM $>$390 \kms and the other (which corresponds to the H$\alpha$ broad component) has FWHM $>$2000 \kms.  In summary, this exercise indicates that there are three kinematical components which we discuss in the following sections.} 

{Velocity field and dispersion maps measured on H$\alpha$ are shown in Figure~\ref{model}. They were obtained from { a fit of a single component for each emission line over the whole FoV, except for the nuclear region where a very broad component was added}. Analysing the velocity field map, we found evidence of rotating gas and that  the velocity field is asymmetric\footnote{This asymmetry can be due to the non-rotational motions of the filamentary structure observed in this galaxy, see section \ref{sec_kine_interp}.}, with the receding side at the south and approaching side at the north. The maximum approaching value is -120~\kms at $\sim$200~pc and the receding one is $\sim$200~\kms at $\sim$100~pc. The line-of-nodes lies in PA$\sim$ 152$\,^{\circ}$.} 

The H$\alpha$ velocity dispersion map { corrected by the instrumental broadening ($\sigma$, Figure~\ref{model}e)}, shows a large range of values { (with a constant mean uncertainty of 4\kms)} of  through the overall FoV, from 100 \kms up to 600 km s$^{-1}$, { the peak of the distribution has an offset of $\sim$ 30~pc to SE with respect to the photometric center, marked with a symbol ``x'' in all maps.
Two different regions can be clearly distinguished from the distribution of $\sigma$ values. The first one is the nuclear region
that shows a $\sigma$ gradient from 300\,\kms to 600 \kms. The second region presents 
a ``dove-like" distribution of $\sigma$ values
between 150 and 300 km s$^{-1}$, with ``wings'' on the 
east and north directions and the ``body'' in the SW direction.}
We noted that, in the rest of the FoV, the velocity field and dispersion velocity have similar values ($\sim$100\,\kms).

\begin{table} 
\caption[Table caption text]{Results for the Gaussian profile fits measured for H$\alpha$ emission line in selected regions.}
\begin{center}
\begin{tabular}{|l|l|l|l|l|} \hline \hline
Region & Set & FWHM (\kms) 	& ${V_{c}}$ (\kms)  & $|\Delta V|$(\kms) \\ \hline
 1 	   & 1   & $215 \pm 28$ &   $-178 \pm 34$ & $196 \pm 50$ \\  
  	   & 2   & $172 \pm 25$ &   $18   \pm 37$ & \\ 
\hline
 2 	   & 1   & $376 \pm 40$ &   $46   \pm 25$ & $407 \pm 37$ \\ 
  	   & 2   & $301 \pm 35$ &   $-361 \pm 28$ &  \\ 
\hline
 3 	   & 1   & $323 \pm 36$ &   $-119 \pm 28$ & $283 \pm 44$ \\ 
  	   & 2   & $194 \pm 27$ &   $-402 \pm 35$ & \\
\hline
 4 	   & 1   & $376 \pm 40$ &   $-23  \pm 25$ & $274 \pm 39$ 	\\ 
  	   & 2   & $280 \pm 33$ &   $-297 \pm 30$ & \\ 
\hline
 Nuc   & Narrow  & 	$1076 \pm 68$  &   $5   \pm 15$  & $558 \pm 17$ \\
	   & Broad   &  $3001 \pm 113$ &   $562  \pm 9$  &  \\ \hline
\hline\hline
\end{tabular}
\end{center}
\label{table:kin}
\end{table}

\section{distribution and ionization mechanisms of the ionized gas}\label{IonG}

\begin{figure*}
%\centering \includegraphics[width=.5\textwidth]{flux_new.eps} 
\centering \includegraphics[width=\textwidth]{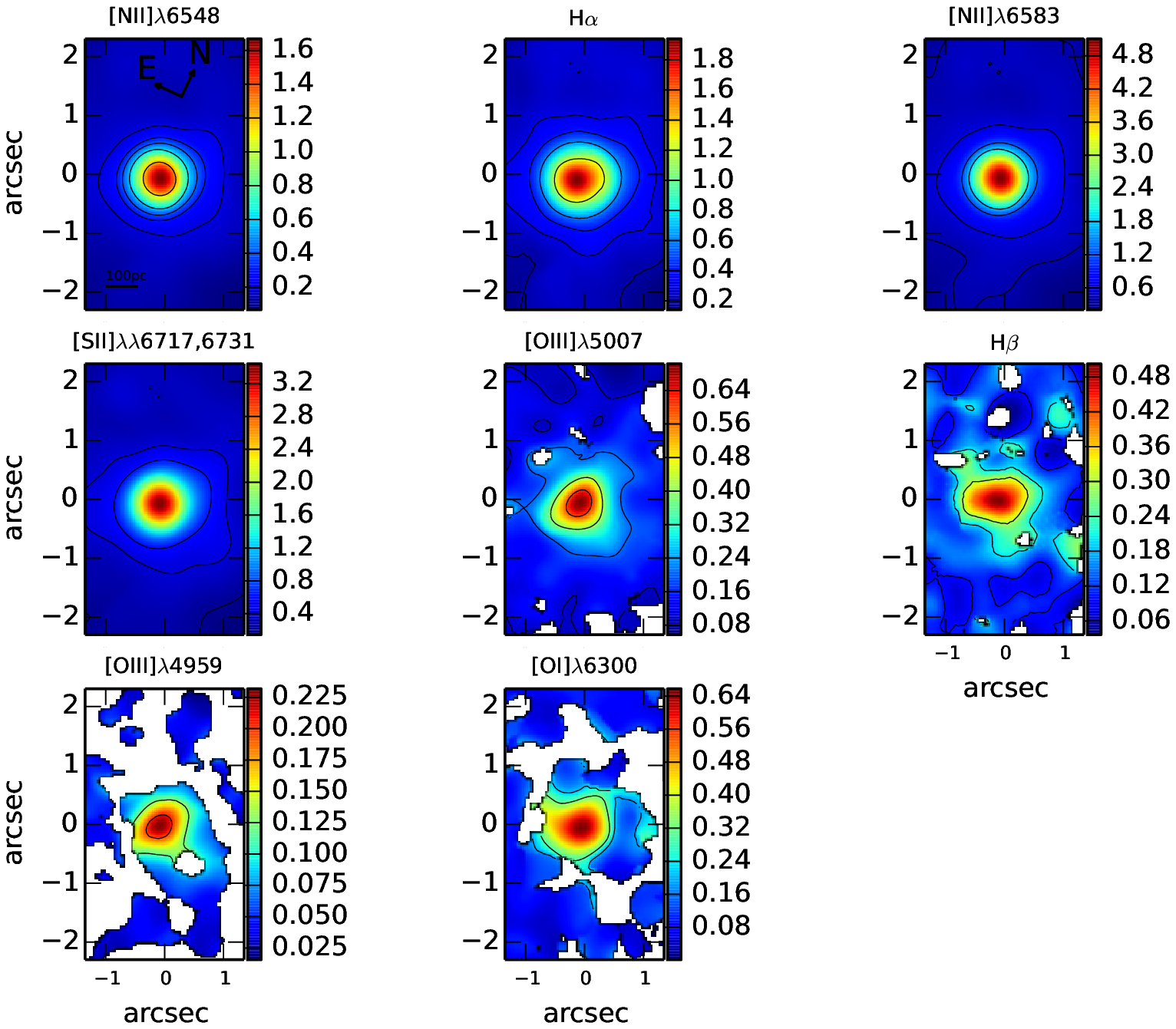} 
\caption{Flux distribution of the emission-lines measured from the spectra free from the stellar contribution. The color bar show the range of flux values expressed in units of 10$^{-15}$ erg cm$^{-2}$ s$^{-1}$.}
\label{flux} 
\end{figure*}

The results of the individual fits obtained from the stellar synthesis can be subtracted spaxel-by-spaxel, letting us with a clean residual data cube, free from the stellar population absorption features (see bottom Figure ~\ref{5044_spec}), thus, allowing us to study the pure emission gas component. { To measure the emission line fluxes we used the same methodology described in Sec.\,\ref{kin}.}

Figure~\ref{flux} present the ionized gas flux distribution for following emission lines:  H$\alpha$ total flux, [NII]$\lambda$6583, [SII]$\lambda$6717+$\lambda$6731 and [OIII]$\lambda$5007. In the inner 1 kpc of the NGC~5044, the gas distribution is irregular. The flux maps for the emission lines [NII] and [SII] show that their emission are more concentrated at the nucleus and indicate a preference of distribution along south-north direction, while the [OIII] map shows emission over the whole FoV, with regions of higher intensity extended along the E-W direction.  
Also, the flux distribution for other emission lines detected in our data (H$\beta$, [OIII]$\lambda$4959 and [OI]$\lambda$6300), do have a lower SNR and are shown for completeness.

{ From the flux maps, we derived the following values for the emission line ratios: 0.1 $<$ log ([NII]$\lambda$6583/H$\alpha$) $<$ 0.5, -0.05 $<$ log [SII]$\lambda\lambda$/H$\alpha$ $<$ 0.5 and -0.4 $<$ log [OIII]$\lambda$5007/H$\beta$ $<$ 0.5, which are typical of LINER-like emission line objects \citep{heckman80,ho08}. }

{ The dominant ionization/excitation mechanism of the ionized gas in NGC~5044 is still controversial. According to \citet{macchetto96} the observed spectra can be explained by post-AGB stars, while \citet{rickes04} suggest that the spectra have contributions of a LLAGN in the nuclear region plus post-AGB stars in the outer parts.}

\begin{figure}
\begin{minipage}[t]{0.45\textwidth}
\includegraphics[width=\textwidth]{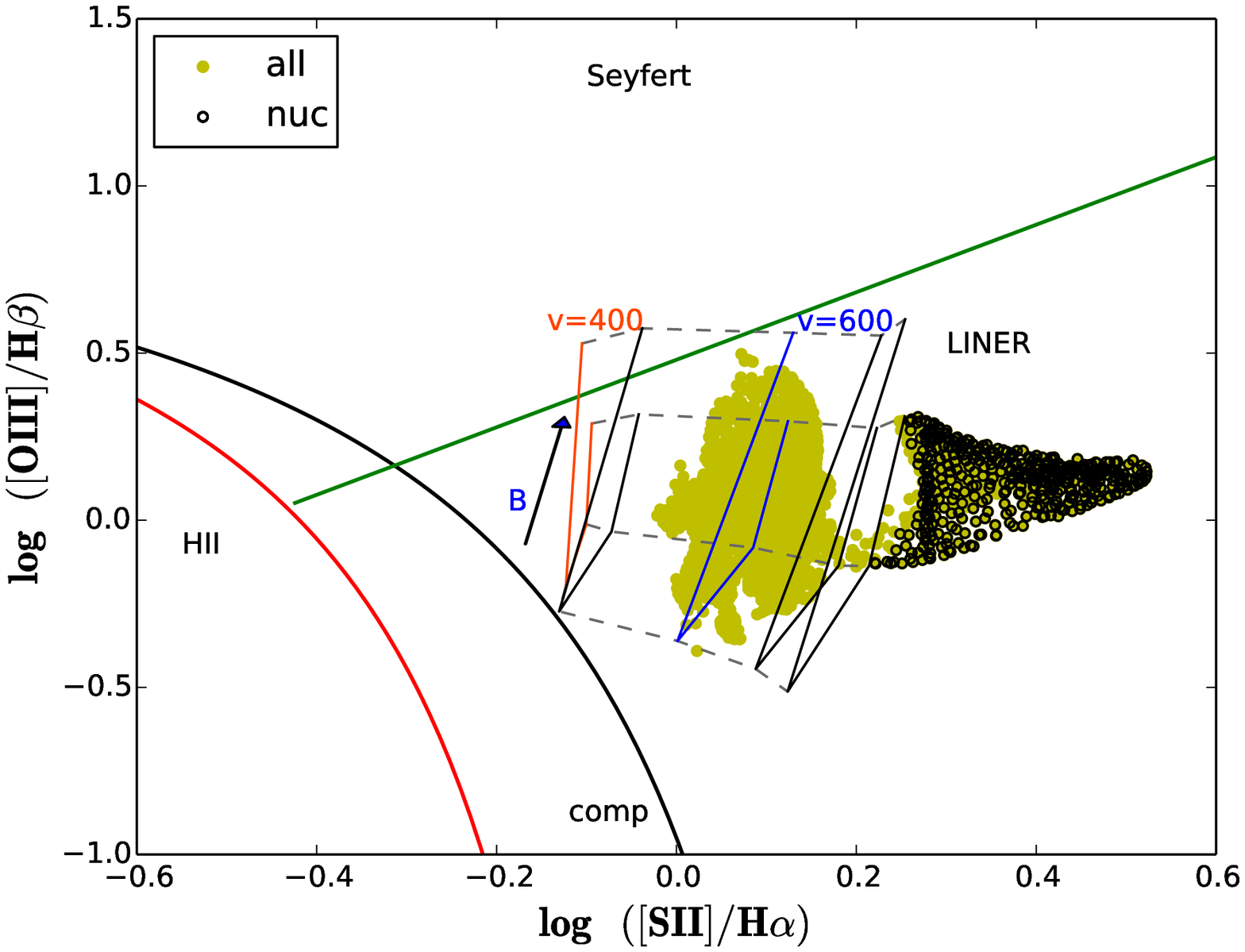}
\end{minipage}
\begin{minipage}[t]{0.45\textwidth}
\includegraphics[width=\textwidth]{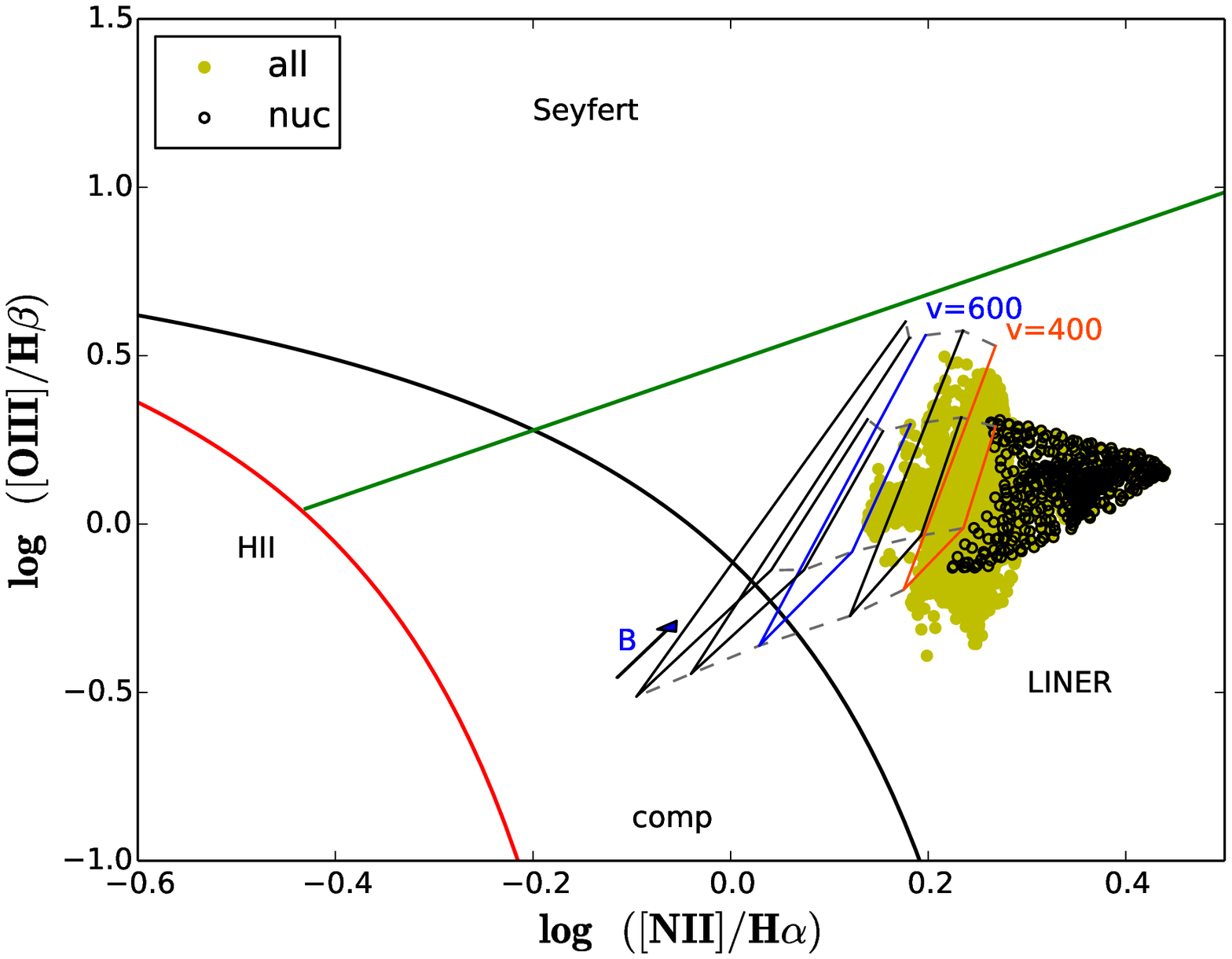}
\end{minipage}
\caption[....]{Diagnostic diagrams. Line ratios of all the spaxels are shown in yellow dots. The nuclear region ($\sim$200~pc) are in black dots. Black (KW01), red (KF03) and green (KW06) lines are the empirically and theoretically derived nuclear activity separations between LINERs, Seyfert and Starbursts. Solid and dashed lines are the grid of the shock models by Allen et al. (2008). The grid has a range in shock velocity (solid line) plotted in steps of 100 \kms, from 400 \kms up to 800 \kms{, we highlighted v=400\,\kms\ (red) and v=600\,\kms\ (blue)}. The magnetic parameter are plotted as the dashed lines with the values 3.23, 0.5, 1 and 2 $\mu$G cm$^{3/2}$.
}
\label{ddNS} 
\end{figure}

{ In order to study the dominant ionization mechanism of the emission line gas in NGC\,5044, we constructed diagnostic diagrams using the emission line ratios mentioned above. The diagnostic diagrams for the nuclear region and for the regions 1 to 4 are shown in Figures~\ref{ddNS} and \ref{dd}, respectively.}

In the diagnostic diagrams, in order to discriminate among the different excitation mechanism, we added the theoretical maximum starburst line of \citet[][hereafter KW01]{kw01} and the theoretical line separating purely star formation galaxies, Seyferts, LINERs and composite galaxies of \citet[][hereafter KW06 line]{kw06}. The empirical line dividing star-forming galaxies and composite objects of \citet[][hereafter KF03]{kf03} was also included.  

Since shocks could explain the LINER-like spectra { \citep{dopita95,heckman80,dopita15}}, we also added to the diagnostic diagrams shock models calculated with the {\sc mappings iii} shock and photoionization code from \citet{allen08}. These shock models include a range of chemical abundance sets, pre-shock densities from 0.01 to 1000 cm$^{-3}$, velocities up to 1000 km/s, magnetic fields from 10$^{-10}$ to 10 $^{-4}$ G, and magnetic parameters (B/$n^{1/2}$) from 10$^{-4}$ to 10 $\mu$G cm$^{3/2}$. { Since, the electron density ($n_e$) of the pre-shock material is an important physical quantity that affects the emission spectrum in radiative shocks, we estimated $n_e$  using [S {\sc II}]$\lambda\lambda$ 6717$\AA$, 6731$\AA$ emission line ratios \citep{osterbrock06,panuzzo11} measured in the regions 1 to 4.  We found 35~$<$~$n_{e}$~$<$200, the lower value corresponds to the region 4, while the higher value was found in the region 3, thus we adopted a mean value of $n_{e}$=100 cm$^{-3}$. The models consistent with our observational data are those of shock without precursor, with solar metallicity and with pre-shock density of $n_e$=100 cm$^{-3}$.}

In Figure~\ref{ddNS} we show the ratios obtained for all the spaxels along the FoV (yellow) and highlight those of the nuclear region (black). It is clear that the nuclear region presents some discrepancy with respect to the circumnuclear one, suggesting a mixing of ionization mechanisms, since the black circles are in the LINER region, while the circumnuclear (yellow) lies in the shock zone. 

{ In addition, we investigate in detail the excitation mechanism of the selected regions 1, 2, 3 and 4 identified in the H$\alpha$  channel maps (Figure~\ref{Channelmaps}). For each one of these regions we constructed diagnostic diagrams involving the line ratios [O {\sc iii}]/H$\beta ~ \times$ [S {\sc ii}]/H$\alpha$ and  [O {\sc iii}]/H$\beta ~ \times$ [N {\sc ii}]/H$\alpha$ (Figure~\ref{dd}). For the diagnostic diagrams using the [SII]/H$\alpha$ ratio (which is more sensitive to shocks), we found shock velocities $>$~550\,km\,s$^{-1}$, while the shock models for [NII]/H$\alpha$ are not able to predict the observed values for regions 1, 3, and 4. However, region 2 can be explained by models with velocities of $\sim$400\,\kms, for the [NII]/H$\alpha$ ratio. The above results suggests that the dominant excitation mechanism in the identified regions is compatible with shock waves with velocity $>$ 550\,\kms (for the [SII]/H$\alpha$ ratio) moving in a environment with density of $n_e$ $=$ 100 cm$^{-3}$. Therefore, in the inner kpc of this galaxy there are two ionization mechanisms taking place: in the center ($\sim$200 pc) the emission lines are dominated by photoionization due to a LLAGN, while in the circumnuclear region the gas emission is can be shock dominated. In addition this result is in agreement with the fact that we are finding a significant fraction of the FC in the central region (see Fig. \ref{5044_spec}) and with the broad component that we found for H$\alpha$ (see Figure.~\ref{espectros}-Nuc), that indicates the presence of an AGN. \\
\\
\\

}

\begin{figure*}
\begin{minipage}[t]{0.43\textwidth}
\includegraphics[width=\textwidth]{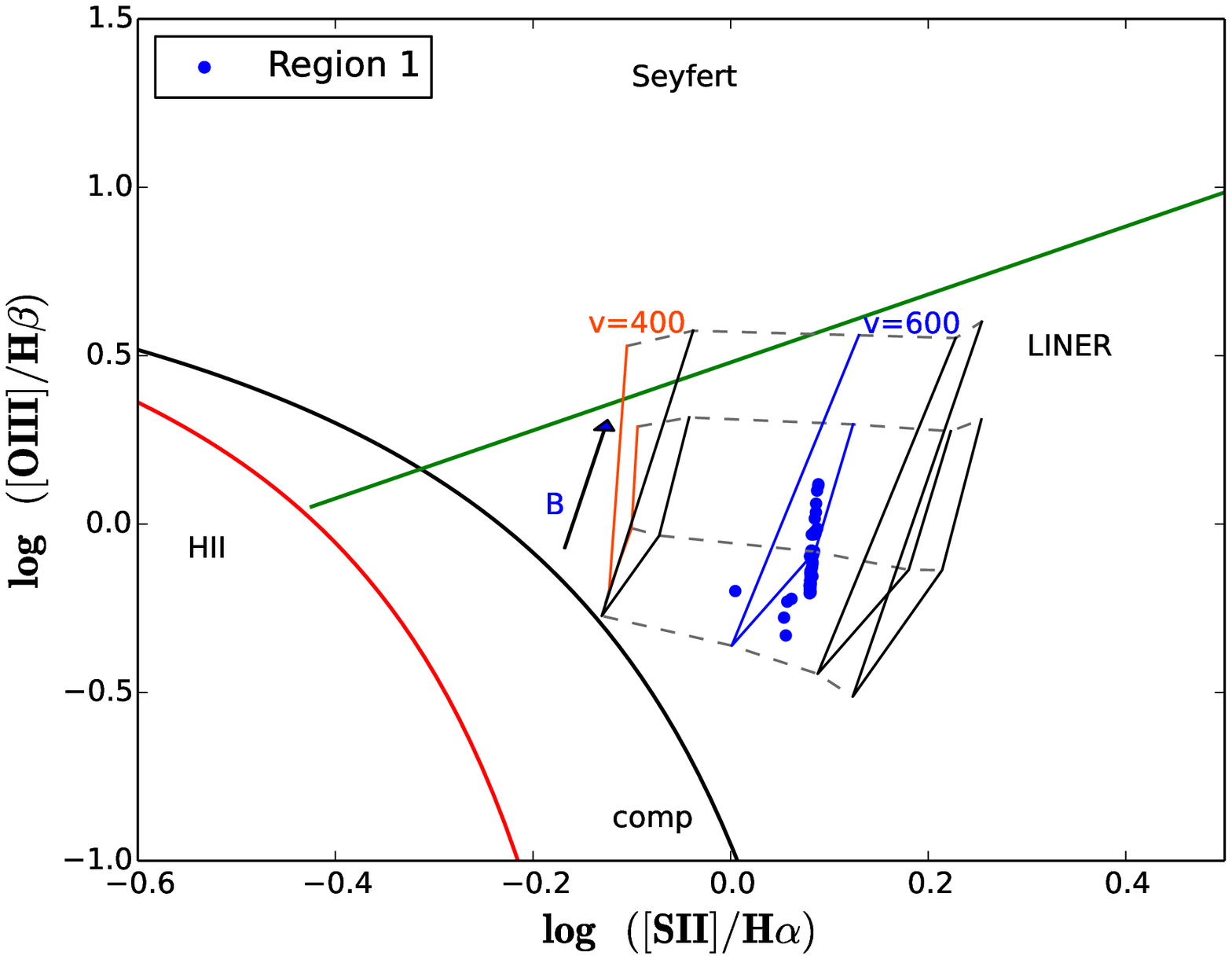}
\end{minipage}
\begin{minipage}[t]{0.43\textwidth}
\includegraphics[width=\textwidth]{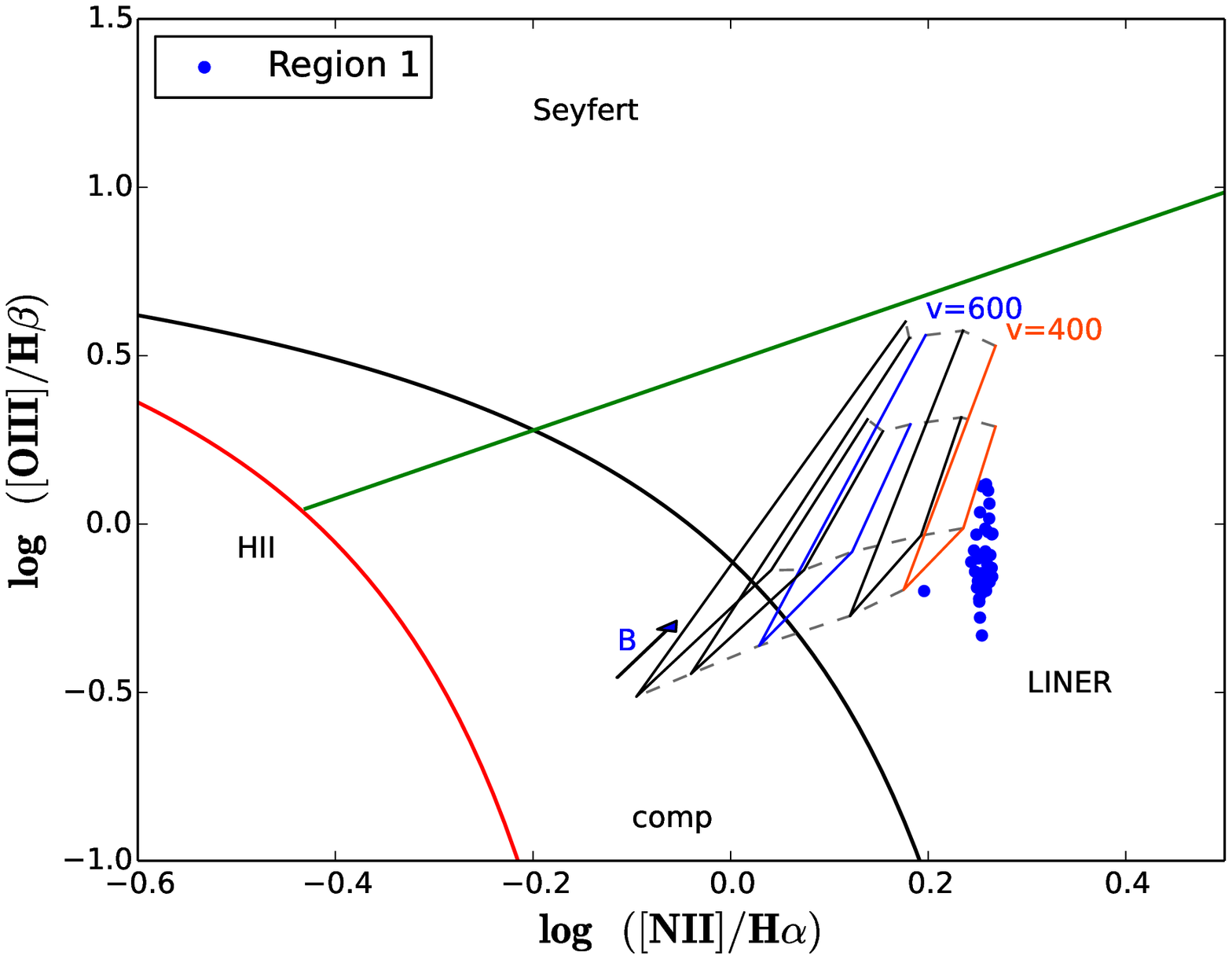}
\end{minipage}
\begin{minipage}[t]{0.43\textwidth}
\includegraphics[width=\textwidth]{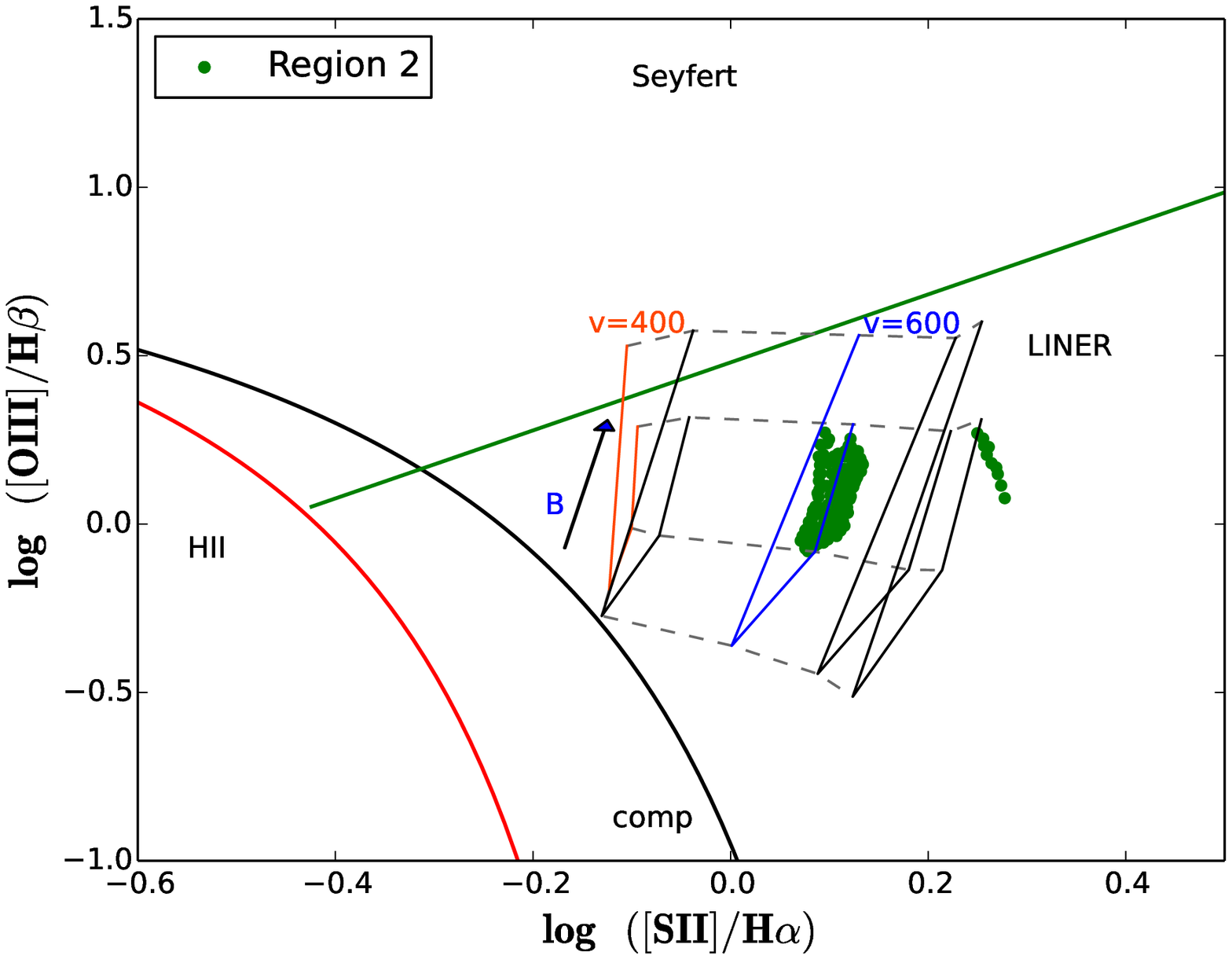}
\end{minipage}
\begin{minipage}[t]{0.43\textwidth}
\includegraphics[width=\textwidth]{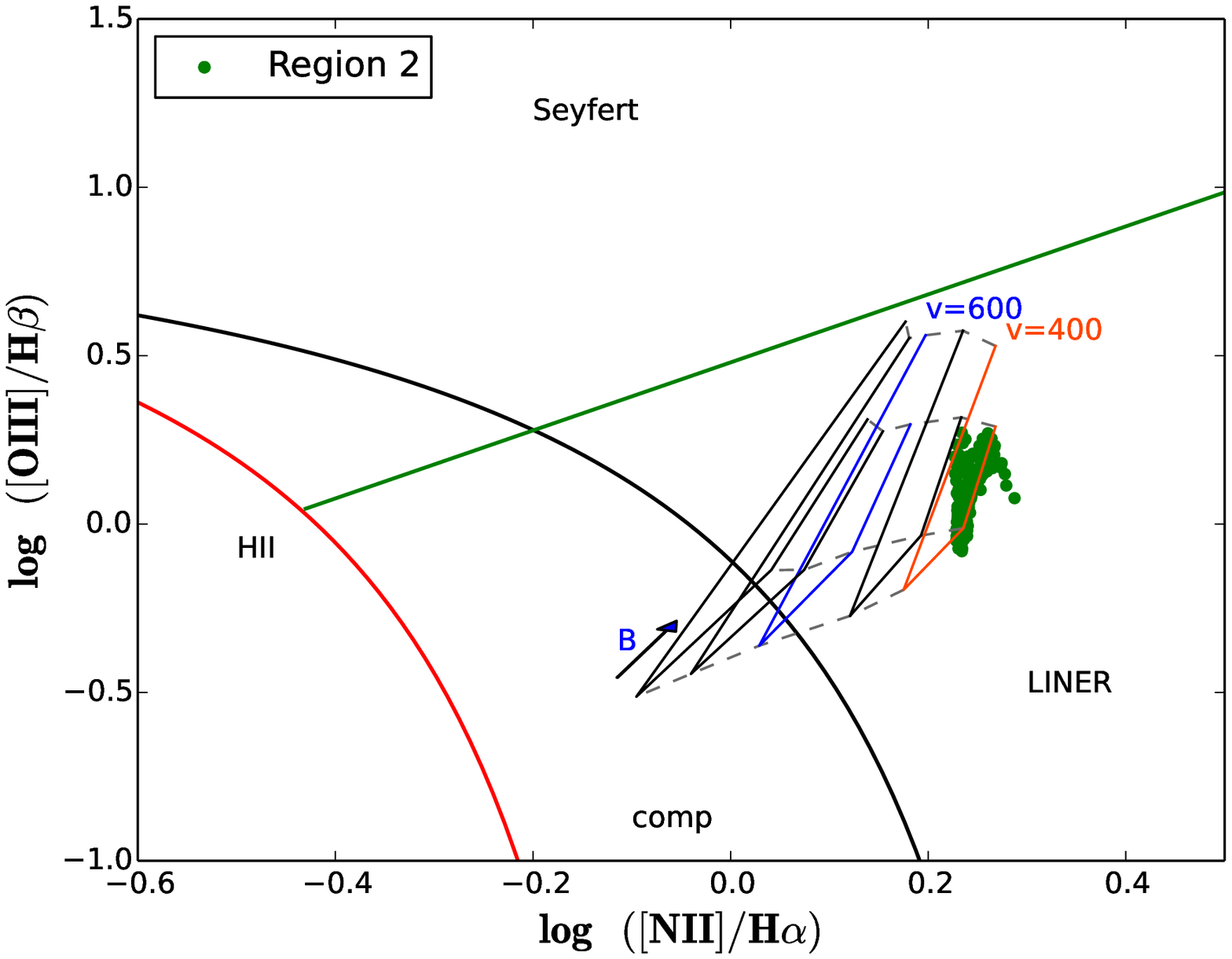}
\end{minipage}
\begin{minipage}[t]{0.43\textwidth}
\includegraphics[width=\textwidth]{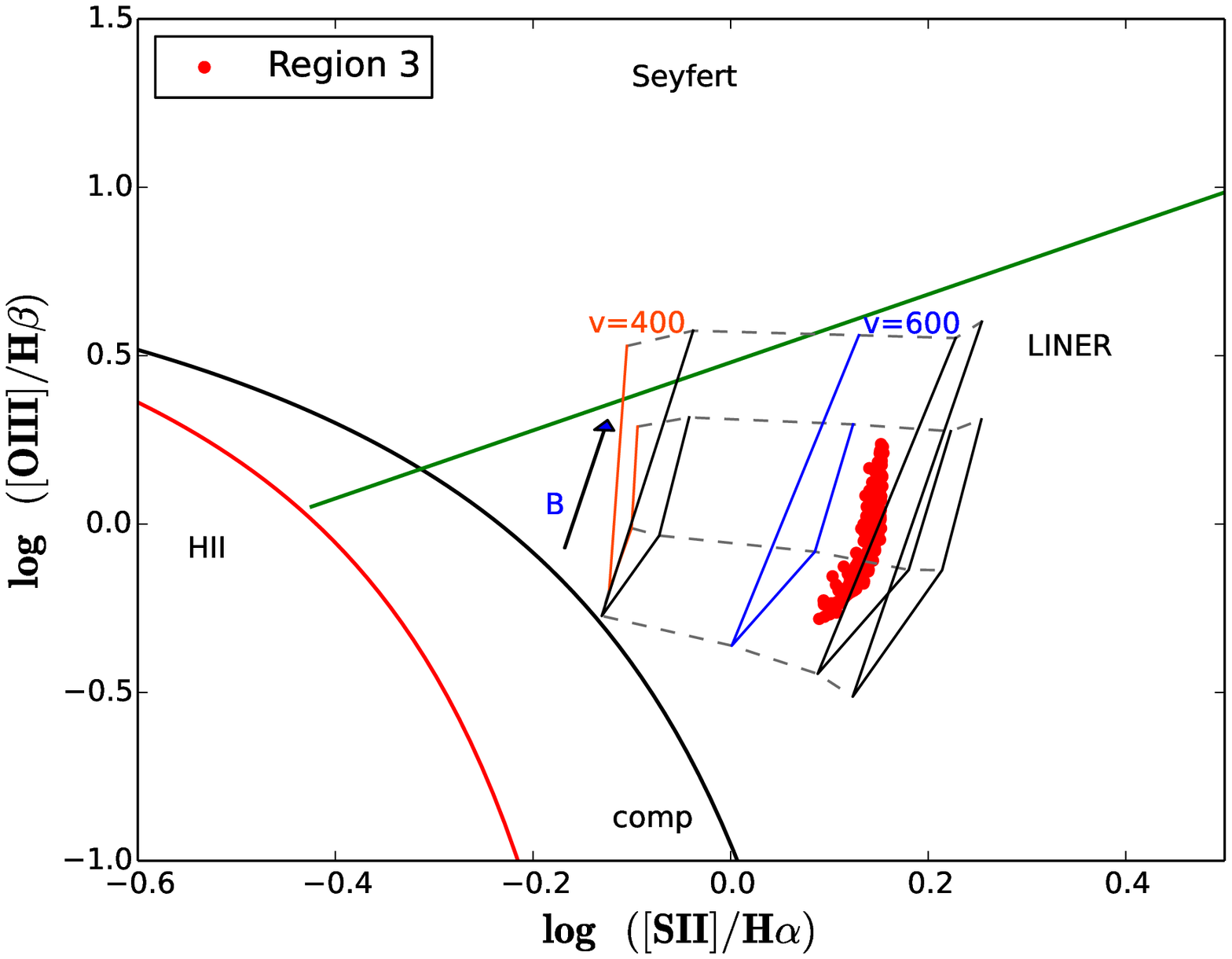}
\end{minipage}
\begin{minipage}[t]{0.43\textwidth}
\includegraphics[width=\textwidth]{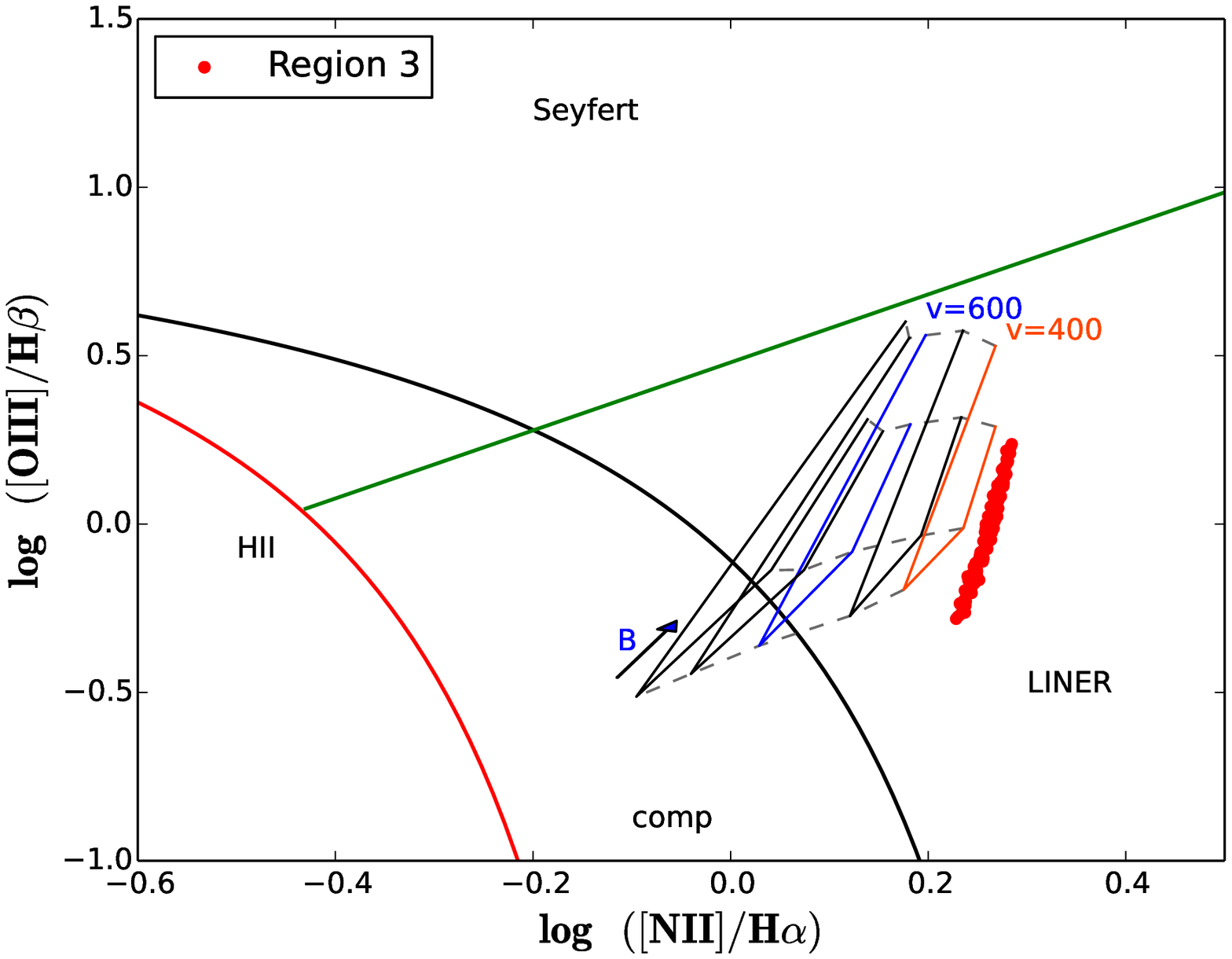}
\end{minipage}
\begin{minipage}[t]{0.43\textwidth}
\includegraphics[width=\textwidth]{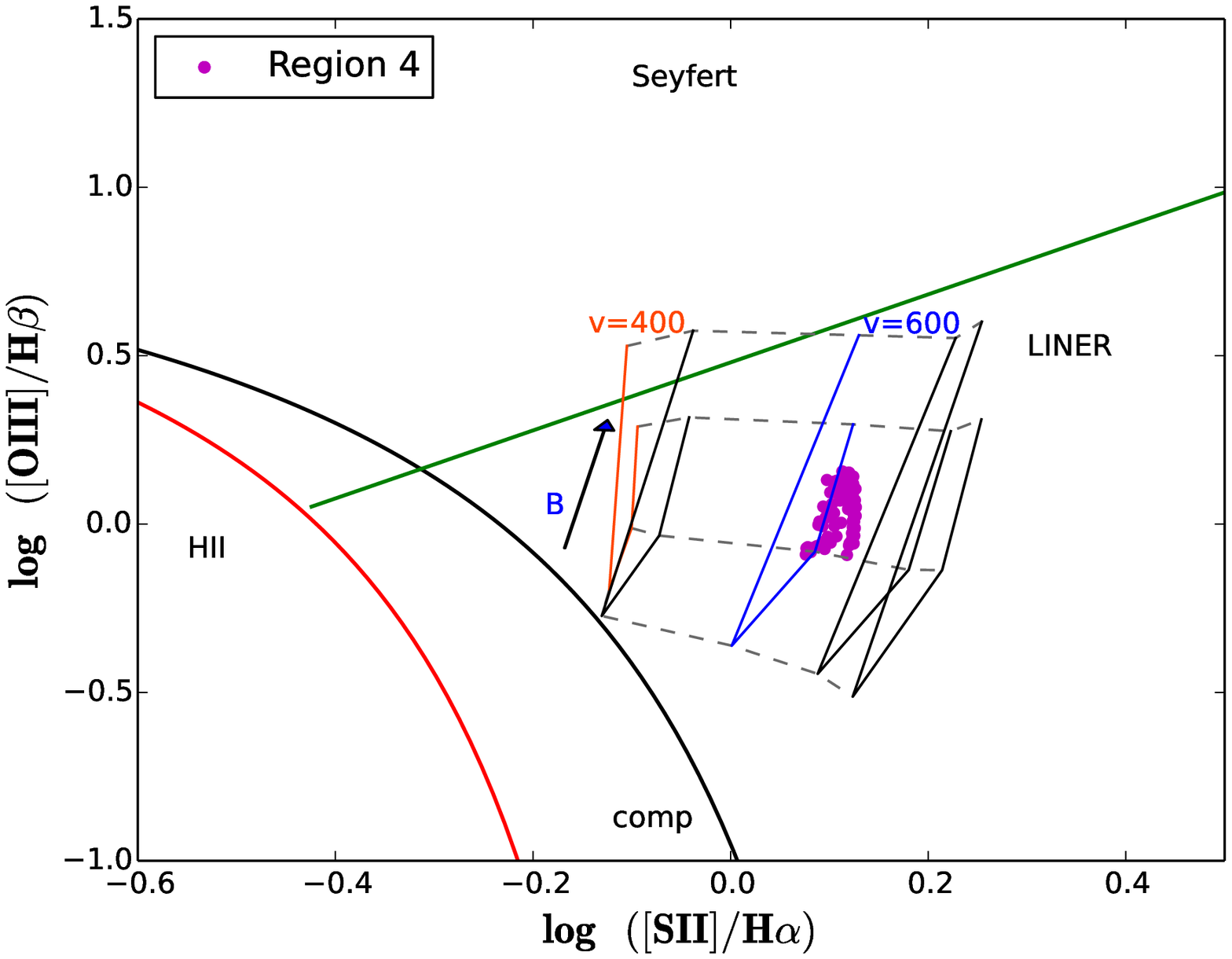}
\end{minipage}
\begin{minipage}[t]{0.43\textwidth}
\includegraphics[width=\textwidth]{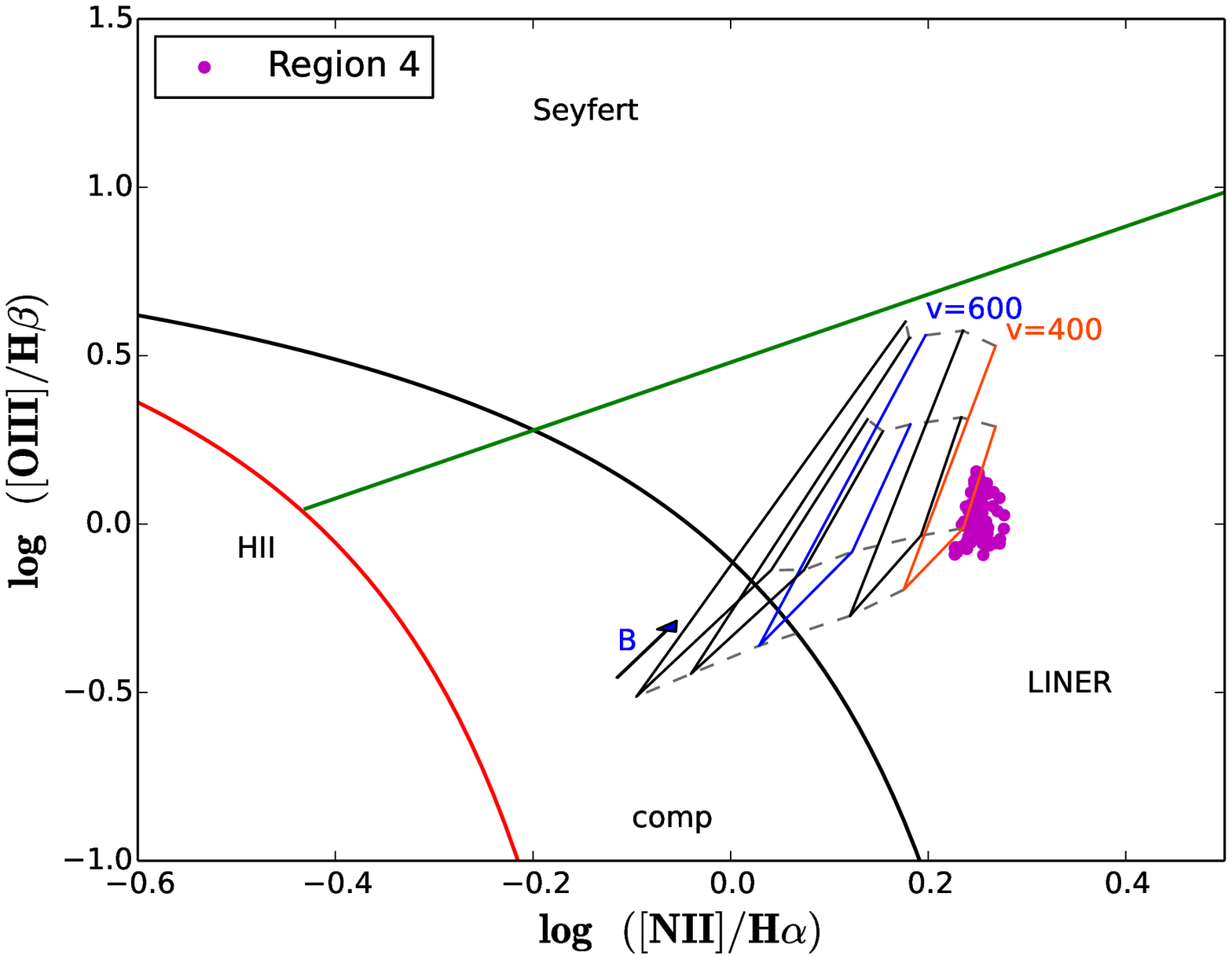}
\end{minipage}
\caption[....]{Same as Figure \ref{ddNS}, but for the regions 1 to 4 selected from the channel maps.}
\label{dd} 
\end{figure*}

\section{The gas velocity field of NGC\,5044: determination of the supermassive black hole mass}
\label{sec_kine_model}

In order to understand the gas kinematics, we have modelled the observed velocity field, 
assuming that the ionized gas follows circular orbits in a plane as proposed by \citet{Bertola91}, summarized by the following equation:

\begin{equation}
{\rm V(R,\psi)= V_{s}+\frac{AR cos(\psi-\psi_{0})sin\theta cos^p \theta}{R^2[sin^2(\psi-\psi_{0})+cos^2 \theta cos^2(\psi-\psi_{0})]+c_{0}^2 cos^2 \theta^{\frac{p}{2}}}}.
\label{eqbertola}
\end{equation}

This results in a velocity curve that increases linearly at small radii and becomes 
proportional to $r^{1 - p}$ for the external regions. $R$ and $\psi$ are the radial and
angular coordinates of a given pixel in the plane of the sky; $A$ is related to the amplitude
of the rotation curve; $c_{0}$ is a concentration 
parameter. The parameter $p$ defines the form of the rotation curve,
varying in the range between 1 (logarithm potential) and 1.5
(Keplerian potential), which is the range of values 
expected for galaxies \citep{Bertola91}; The parameter V$_{s}$ is the 
systemic velocity; $\psi_{0}$ is the position angle of the line of 
nodes; { $\theta$} is the disk inclination ($\theta$ = 0 for face-on 
disks); finally,  there are two implicit parameters: the coordinates of the
kinematic center, $R_{cx}$ and $R_{cy}$. { The equation \ref{eqbertola} has a 
degeneration between $A$ and $\theta$ parameters, which makes not reliable the $\theta$ value derived to 
the fit. Then, we did an estimative for $\theta$ of 42$\degr\pm2\degr$ from the 
outer isophotes of the H$\alpha$ flux image. Thus, this parameter was fixed for the fit.}

\begin{figure*}
\centering \includegraphics[width=\textwidth]{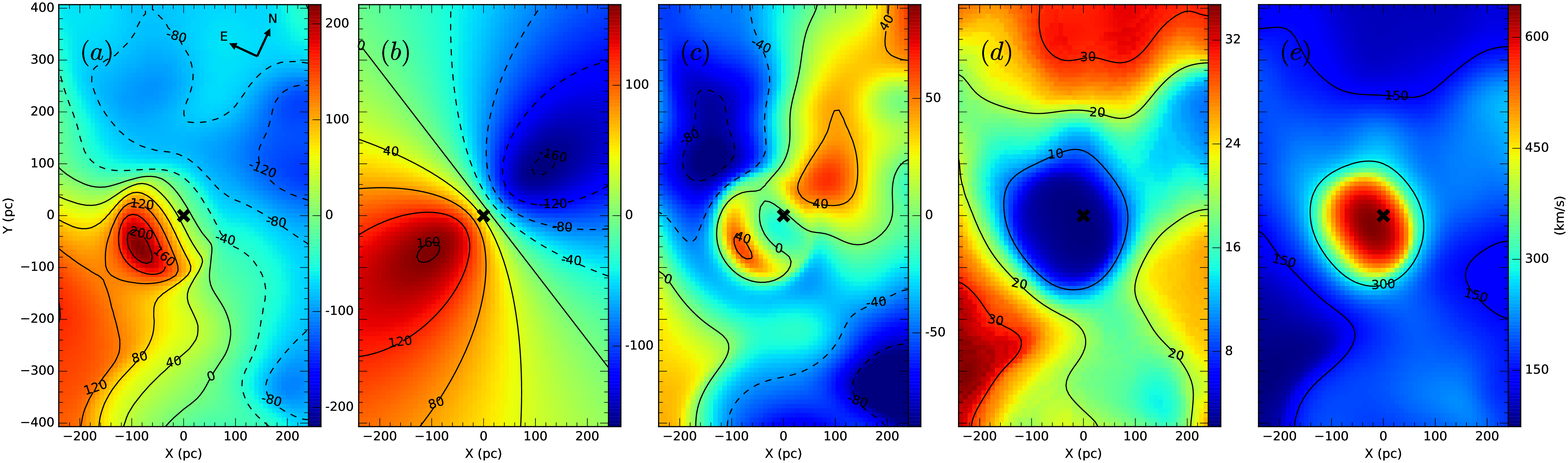}
\caption{From left to right the panels shows: $(a)$. the velocity field map with iso-velocities, $(b)$ the best model fitted; $(c)$ the residual map of velocities, $(d)$ the error map of velocity field $\delta_{v_r}$ and $(e)$ the velocity dispersion ($\sigma$) map. The photometric center of the galaxy is marked with a symbol ``x'' in all maps, and it match well the kinematic center (panel $(b)$), while the peak of the velocity dispersion is at 30 pc to SE.}
\label{model} 
\end{figure*}

 { We used a Levenberg-Marquardt method to fit 
the observed velocity field. The weighted fit 
was done considering the error map of  $\delta_{v_r}$ (Fig. \ref{model}d). The fitted parameters and their respectively errors} are given in Table
~\ref{table:disk}. The $R_{cx}$ and $R_{cy}$ are not listed in this Table,
instead of them, we give the difference between kinematic and photometric centres, in the sky plane, $\Delta\,x$ and $\Delta\,y$. { The 
reduce $\chi^2$ of the fit is 10.8, this high value is due to 
the perturbations of the observed velocity field, and is reflected in the residual map (Fig. \ref{model}c), which presents non-rotational strong kinematic components, 
they are discussed and analysed in detail in Sec. \ref{sec_kine_interp}.}

\begin{table} 
\caption[]{Gas velocity field parameters.}
\begin{center}
\begin{tabular}{|l|c|} \hline \hline
Parameter             & Values \\ \hline
A (km s$^{-1}$)       & $1285\pm189$\\  
$c_{0}$ (pc)          & $95\pm5$     \\  
p                     & $1.5\pm0.1$   \\ 
v$_{s}$ (km s$^{-1}$) & $2782\pm1$ \\ 
$\psi_{0}$            & $152\degr\pm2\degr$ \\ 
$\Delta x$  (pc)      & $0\pm1$   \\ 
$\Delta y$  (pc)      & $18\pm1$  \\ \hline
\end{tabular}
\end{center}
\label{table:disk}
\end{table}

The galaxy kinematic centre has a small offset, of { 18 pc to NE} 
direction, with respect to its photometric centre. 
Considering that our angular resolution is { $\sim$ 125 pc}
this offset is negligible and we conclude that the kinematic centre
coincides (within the uncertainties) with the photometric nucleus.

 The rotating disk model that provided the best fit of the observed velocities is shown in Fig.~\ref{model}b. The rotation curve derived from this model is presented in Figure~\ref{curva}. It is clear that the gas reaches its maximum rotational velocity ( 240 km s$^{-1}$) at a distance of 136 pc. In addition, the derived value for the $p$ parameter { (p$\sim$1.5)} indicates that the gas distribution has a Keplerian behaviour and thus the dynamic in this region must be dominated by a high mass concentration.  Assuming the Keplerian approximation for the gas motion ($M=r \times V_{{\rm rot}}^{2}/G$), { we estimate a mass of $1.9\pm0.9\times10^9M_{\odot}$ within the inner 136\,pc.

The value of this mass concentration can be verified using the M-$\sigma$ relation \citep{kormendy13} as follows:

\begin{equation}
M_{bh} = 0.309^{+0.037}_{-0.033}\times10^{9} M_{\odot}\left(\frac{\sigma_\star}{\rm 200 km s^{-1}} \right)^{4.38\pm0.29},
\end{equation}
} 
where $\sigma_\star$ is the stellar velocity dispersion measured in an
aperture with radius of $R_e$/8, with $R_e$ being the half-light radius. 
We derived this radius for NGC~5044  from its brightness surface profile, using a $K_{s}$ band image
taken from 2MASS Extended Source Image Catalogue \citep{skrutskie06}. The obtained value is $R_{e}$=29\arcsec, which corresponds to 5.1\,kpc, the required value of $R_e$/8 is $\sim$ 600\,pc, that is close to the datacube FoV.  { The $\sigma_\star$  was determined from the absorption lines  Mgb$\lambda$5167, FeI$\lambda$5270 and FeI$\lambda$5335  measured in a single integrated spectrum obtained collapsing all the non stellar subtracted cube spectra, since we are only interested to measure
the FWHM of these stellar lines}. The FWHM of the lines was corrected for instrumental broadening. The resulting $\sigma_\star$ is the mean weighted by the flux of each absorption line, and the uncertainties was estimated from the standard deviation of these measurements. Then, the found value of $\sigma_\star$ for NGC~5044 is { $300\pm57$ \kms}. This value is in agreement with the 250\,\kms found for the galaxy center in \citet{caon00}.} { The $M_{bh}$ derived using this $\sigma_\star$ is  $1.8\pm1.6\times10^{9}\,M_{\odot}$, which is in full agreement with the value obtained with the Keplerian approximation.

With the purpose of testing the possibility that the peculiarities in  the velocity field obtained in Sec.~4 [e.g. a steep rise toward the center (300 - 600 \kms), and the rotation curve reaching its maximum value at a small radius (136~pc)]} are associated with the presence of a black hole, we plotted in Figure~\ref{curva} the Keplerian rotation curve  expected for a SMBH with { M=1.8$\pm1.6\times10^{9}\,M_{\odot}$} (dash-dotted line), the observed {(red line)} and modelled {(black line)} rotation curves for the central region of the galaxy. We also include the rotation curve derived from the S\'{e}rsic profile {(dashed line)} using the K band brightness surface profile \citep[see][for details on methodology]{hernandez13,hernandez15}. From Figure~\ref{curva}, it is clear that { for the inner 136~pc (dashed vertical line),} the central kinematics is dominated by the SMBH potential. This would explain why the maximum rotation curve of the gas and the steep rise in the dispersion velocity profile is observed inside this small radius. Otherwise, without the presence of a SMBH the observed rotation curve would follows that one derived from the S\'{e}rsic profile. A similar kinematic behaviors as that we found for NGC~5044 have been observed in other early type galaxies, for example the elliptical M~87 that harbor a SMBH \citep[3$\times10^{9}\,M_{\odot}$][]{macchetto97,ho00} and the lenticular galaxy NGC~3115 \citep[$10^{9}\,M_{\odot}$][]{kormendy00}. 
Taking the evidences presented here, we conclude that the nuclear gas motions are driven by the gravitational potential of the central SMBH, { with mass of $1.8\pm1.6\times10^{9}\,M_{\odot}$. }

\begin{figure}
\centering \includegraphics[width=\columnwidth]{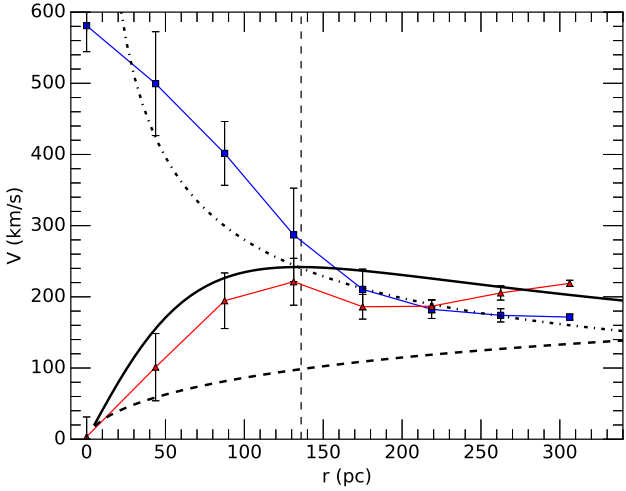}
\caption{The projected observed rotation curve (red line), overlaid the 
modelled one (black). The blue line (square points) is the mean velocity 
dispersion profile along of line-of-nodes. The dash-dotted line is the 
BH Keplerian rotation curve. The dashed line is the rotation curve derived 
from S\'{e}rsic's  K brightness profile. Note that the gas dynamics on the left side of the 
vertical line { (at 136 pc)} is dominated by the BH potential.}
\label{curva} 
\end{figure}

\section{Evidence of gas inflow in NGC~5044}
\label{sec_kine_interp}

In order to derive the non-rotational velocities, resulting
from radial or perpendicular motions, we used
the residual velocity map obtained by subtracting the velocity field model from the 
observed one (Fig.~\ref{model}c). 
We can distinguish in the residual map two blueshifted regions 
with velocities around $\sim$ $-40$\,\kms, one at east and other at 
west. { Both regions show a peak of $-80$\,\kms.} 
On the other hand, two redshifted structures are clearly observed 
along the galaxy's line-of-nodes (north and south), with velocities of 
40\,\kms. These regions 
have the velocity dispersion of 150 \kms (see Figure\ref{model}e). 
The detection of these structures is reliable, because their velocities 
are higher than the uncertainties ($\delta_{v_{r}}$) { as seen 
in the velocity field uncertainty map included in Fig.~\ref{model} (panel d).}

\begin{figure*}
%\begin{minipage}[t]{0.27\textwidth}
\begin{minipage}[t]{0.33\textwidth}
\includegraphics[width=\textwidth]{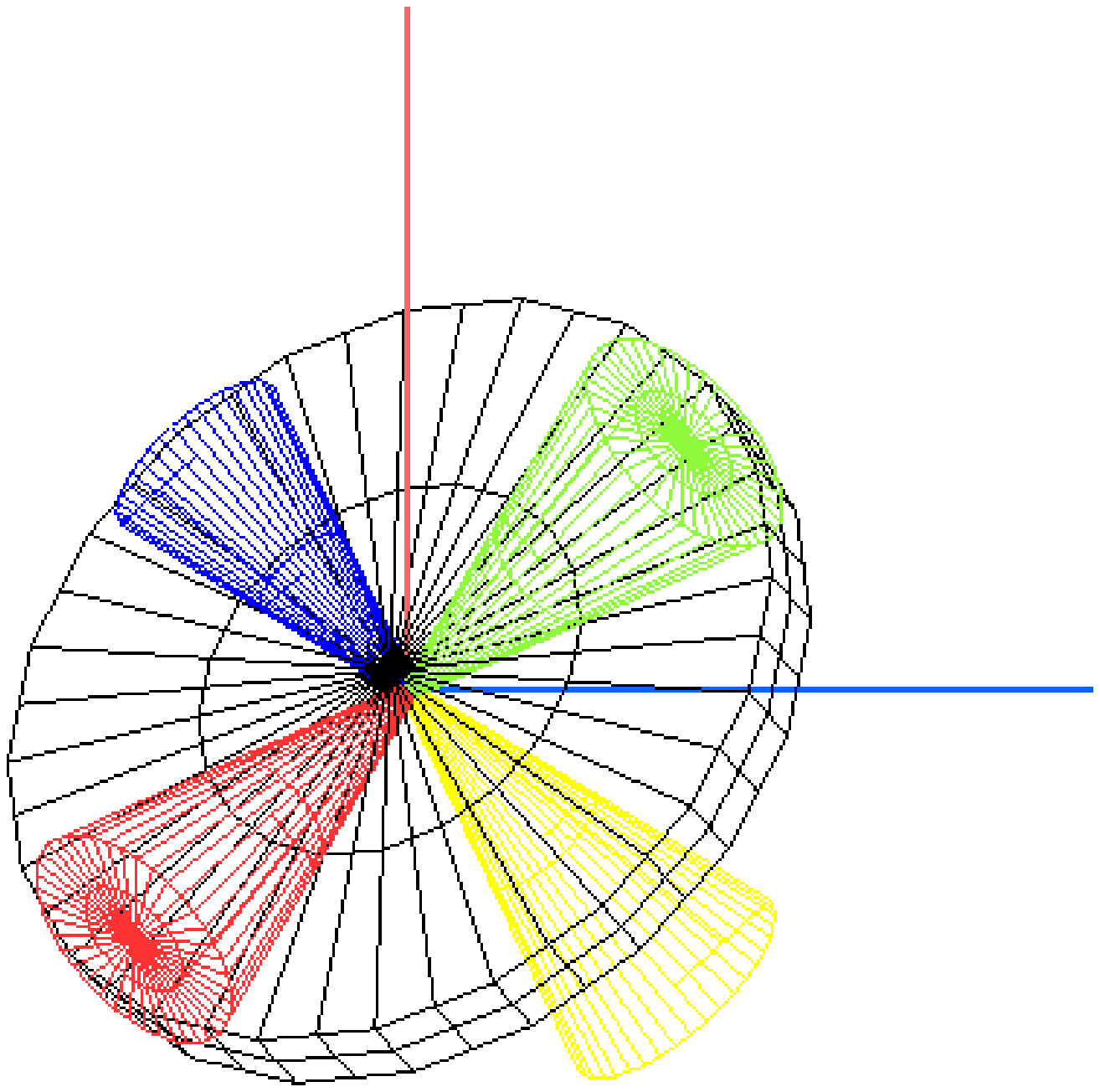}
\end{minipage}
\begin{minipage}[t]{0.3\textwidth}
\includegraphics[width=\textwidth]{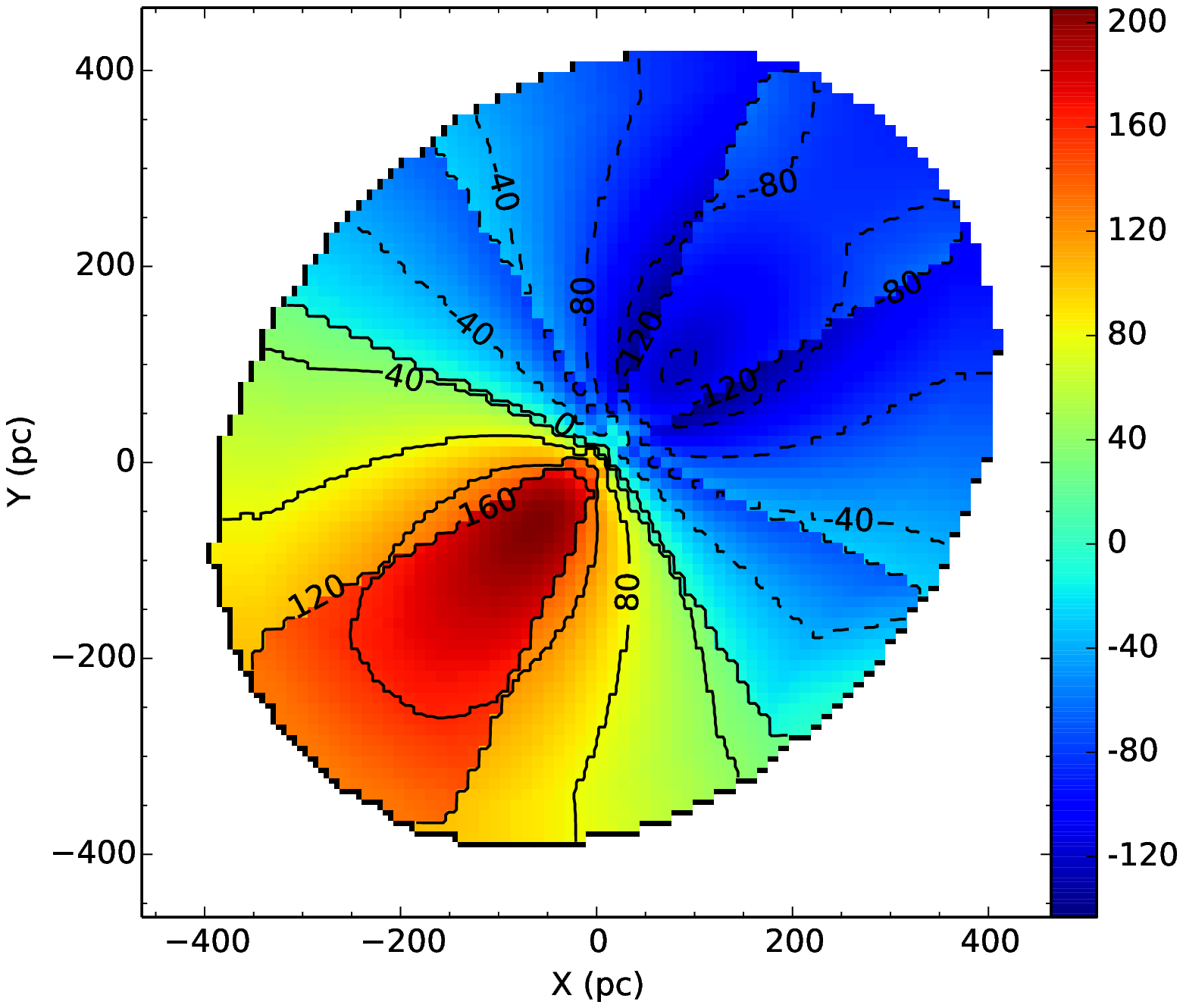}
\end{minipage}
\begin{minipage}[t]{0.3\textwidth}
\includegraphics[width=\textwidth]{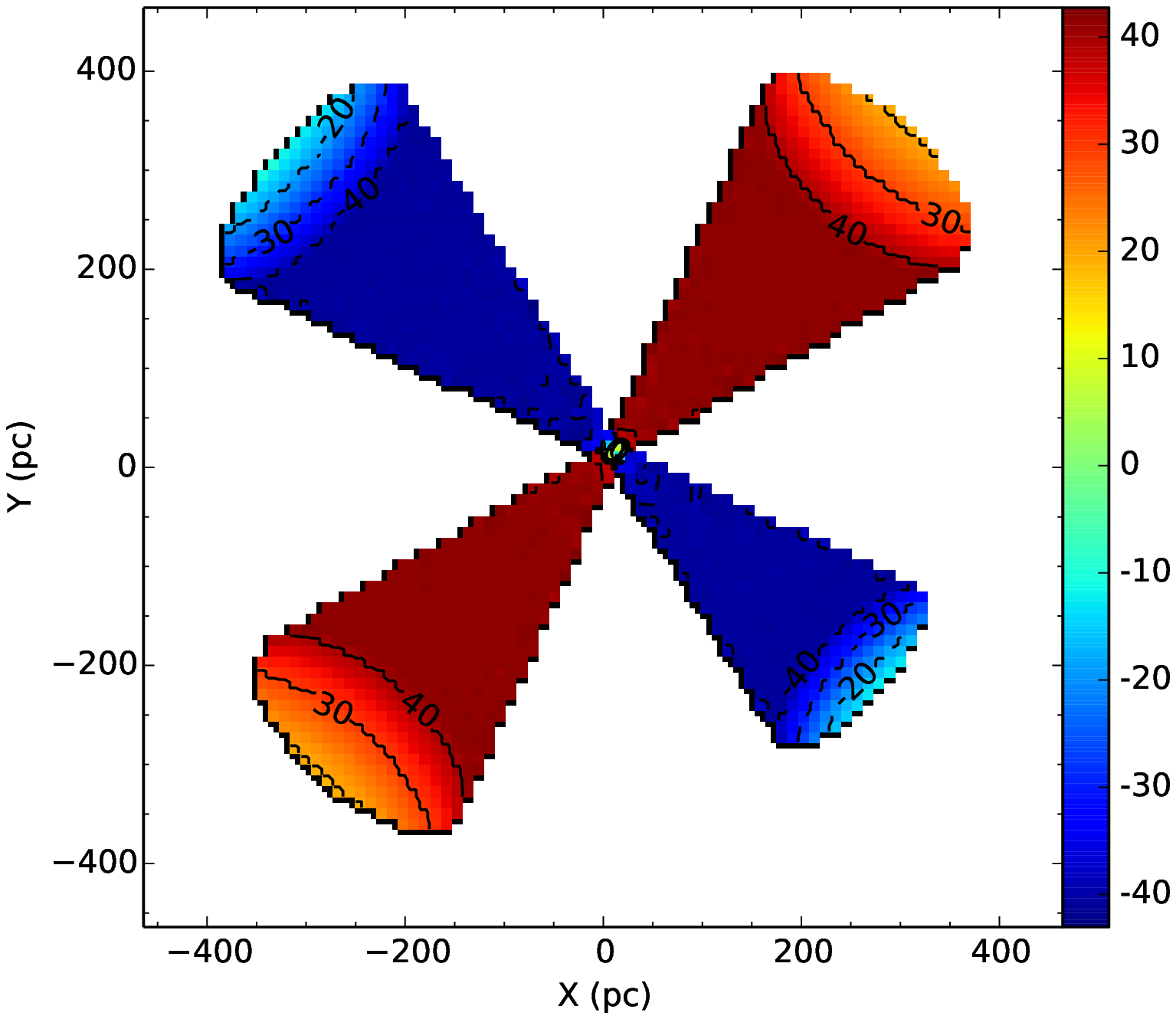}
\end{minipage}
\caption[...]{{ Left panel:} representation in 3D to the arrangement of 
filamentary structures { observed from the line of sight.} The green cone 
represents the north filament, the red represents the south one, the blue 
cone represents the filament at east and the yellow cone represents the 
filament at west (coming from the far side of the galaxy).{ Middle 
panel: the synthetic model of the velocity field along the line of sight. 
Right panel:  the projected velocities of the cones along the line of sight.}}
\label{shape} 
\end{figure*}

With the purpose of interpreting { the structures described above,} we used the Shape 
software \citep{steffen10}, which allows the user to create interactively 3D spatial 
structures with their respective velocity fields. The outcome is a projected 
synthetic velocity field image with a similar orientation to the observer. 
The output velocities are integrated along the line of
sight. We presents in Figure~\ref{shape} the 3D model for NGC~5044. We assume a rotating disk 
represented by a cylinder, whose height is one tenth of the radius 
{ and its} velocity field { is that} obtained in Section \ref{sec_kine_model}, 
whereas we assume for  { all} gas filamentary structures  
{ a geometry of cone with an aperture of $15\degr$, where the 
gas inside it has a velocity vector oriented} toward the center.  { The orientations of the 
cones and the magnitudes of their respective velocity vectors were} constrained requiring that 
{ projected velocities} of these structure, along of the line of sight, match  { the} 
residual map. 
We found that the best orientation for { the cones}  is when: 
a)  The north and south cone-filaments are inclined in 20$\degr$ with respect to 
the disk and they have a PA with respect to the line-of-node of 0$\degr$ and 
{ 180$\degr$}, respectively, { and the magnitudes of their respective 
velocity vectors are $80$\,\kms} b) the cone associated with the east filament 
is inclined { 26$\degr$} with respect to the disk, has a PA of  
{ 90$\degr$}, and { the magnitude of its velocity vector is 
$105$\,\kms}; c) the cone that represents the west filament is 
inclined in { $-62\degr$} with respect to the disk, has a PA of 270$\degr$, and
{ the magnitude of its velocity vector is $105$\,\kms}.
{ The resulting geometrical configuration, along the line of sight, to 
represent the gas filamentary structures is shown in Fig. \ref{shape}a.}
The synthetic model of the velocity field is presented in in Fig. \ref{shape}b. 
It is created from the sum of the disk and the filamentary structures models, 
which would represent the observed velocity field as seen in Fig. \ref{model}a. 
We can see that the model is able to recovery the asymmetry in the observed 
velocity field. In addition, we present in Fig.  \ref{shape}c the projected 
velocities of the cones along the line of sight. It reproduces well the 
values of residual velocity map (Fig.  \ref{model}c). This suggest that a suitable 
interpretation of the residual map could be that the gas in the 
filamentary structures is falling toward the galaxy center.
{ This result is in agreement with the findings of \citet{david14}, 
which  suggested that some of the  
molecular CO clouds they detect are infalling towards 
the galaxy in a ballistic trajectory.  
However, we should be caution with our results since they are a simple model 
approach to reproduce the complex velocity field of the inner 600\,pc of NGC5044. }

{ Using the above results we are able to make an estimative of the inflow rate of the gas in the filamentary 
structure from their observed H$\alpha$ luminosity and the mean magnitude of the 
velocity of the cone-filaments. Thus, first we integrated the H$\alpha$ flux in all FoV and found 
${F_{H\alpha}}$= 1.65$\times$10$^{-12}$\,erg\,cm$^{-2}$\,s$^{-1}$, which at the galaxy distance (Table~\ref{table:prop}) results in ${L_{H\alpha}}$=2.7$\times$10$^{41}$ erg s$^{-1}$. Note that the value obtained here is higher than that found by \citet{macchetto96} integrated over the entire galaxy. We attribute this difference to indirect calibration approach used by \citet{macchetto96} . 

The ionized gas mass, $M_{HII}$, can be estimated as follows: 

\begin{equation}                                                                                                                                    
\centering
M_{HII} = \frac{L_{H\alpha}}{h\nu_{H\alpha}}\frac{m_p}{n_e\alpha^{eff}_{H\alpha}},
\end{equation} 
where, $m_p$ is the proton mass and $\alpha^{eff}_{H\alpha}$ is the effective recombination 
coefficient for H$\alpha$. Then, by assuming the ``case B'' for the recombination \citep{osterbrock06}, 
adopting a T$_e$=10000\,K and the mean n$_e$= 100\,cm$^{-3}$ around of FoV, we found a
 ${M_{HII}}$=6.38$\times$10$^{6}M_{\odot}$. Finally, the inflow rate is calculated with the following expression:

 \begin{equation}                                                                                                                                    
\centering
\dot M_{inflow} = 4N_{HII} \upsilon A,
\end{equation} 
where $N_{HII}$ is the density of ionized gas, $\upsilon$ the velocity of the inflowing material, 
$A$ is the area of the cone base, and the factor 4 is due to the number of cones.
$N_{HII}$ was estimated adopting a volume of 0.2\,kpc$^3$, which was calculated from the 
equivalent radius of the physical area of the FoV, $r_{FoV}=0.41\,kpc^2$. 
$\upsilon=93$\,km\,s$^{-1}$ is the mean value of all cones. $A=0.03\,kpc^2$ was calculated 
from the $r_{FoV}$ and aperture angle of the cones ($15\degr$). Then, we found a 
value of $\dot M_{inflow}\approx 0.4\, M_{\sun}\,yr^{-1}$.    
We can now compare it to the mass accretion rate necessary to power the active nucleus, by using
 \begin{equation}                                                                                                                                    
\centering
\dot M = L_{bol}/(\eta c^2),
\end{equation} 
where $L_{bol}$ is the bolometric luminosity, $\eta$ is the efficiency 
of conversion of the rest-mass energy of the accreted material into radiation,
and $c$ is the light speed. We approximated the bolometric luminosity as 
$L_{bol}\approx100L_{H\alpha}\approx2.7\times10^{43}$\,erg\,s$^{-1}$ (e.g., Ho 1999, 
et al. 2001). We adopted a value of 0.1 for $\eta$, which is a typical
value for a  optically thick and geometrically thin  accretion
disc (e.g. Frank, King \& Raine 2002). Then, we obtain an accretion 
rate of $\dot M\approx 0.005\, M_{\sun}\,yr^{-1}$. This value is a factor of $10^2$ lower than 
the inflow rate calculated. 

One question which rises now is: are these filamentary structures an extension of the larger-scale (kpc) filaments earlier studied by \citet{macchetto96}? \citet{temi07} found a remarkably spatial correlation between the filamentary structures in H$\alpha$ and dust detected at 8 and 70 $\mu$m. These authors claimed that the dust present in the filaments most likely comes from the galaxy centre. Based on these results, they proposed that NGC~5044 has suffered a recently (10$^7$ yr ago) feedback where the dust located in the nuclear region was disrupted, heated and transported outward to several kiloparsecs. Evidences of bubbles were found by \citet{gastaldello09} which reported that the inner 10\,kpc of 
this galaxy harbors a pair of cavities together with a set of bright X-ray filaments coincident with H$\alpha$ and dust emission. In addition, \citet{gastaldello09} 
make an estimate of the release time for the bubbles as 1.2$\times$10$^7$ yr, supporting 
the scenario proposed by \citet{temi07} for this galaxy. Besides, our results point to ionized gas flowing towards the galaxy center, which one can associate with the kpc scale filamentary structure reported for this galaxy. In addition, we found strong evidence for the presence of an accreting SMBH in the nuclear region of NGC\,5044, besides a broad component for H$\alpha$ therefore, we speculate that this gas inflow would be originated and transported from large kpc scale to the nucleus of NGC~5044 through the filamentary structures and part of it being accreted by the SMBH and making NCG\,5044 active. 
This would be in agreement with the scenario where the cooling flows into central galaxy of galaxy clusters (which is the case of NGC\,5044) are regulated by cyclic feedback episodes \cite[e.g.][and references therein]{peterson06}. }

\section{CONCLUSIONS}

We present here, for the first time, high spatial resolution Gemini GMOS/IFU data of the elliptical galaxy NGC~5044 to map the stellar population, emission line fluxes distribution and gas kinematics in the inner kpc, with an spatial resolution of 125\,pc. Our mains conclusions are:

\begin{itemize}
{ 
\item The continuum emission of the inner kpc of NGC\,5044 is dominated by an old stellar population. We also detect an intermediate age population ($\sim$ 40\% of an 900 Myr SSP) surrounding the galaxy nucleus. A featureless contribution of $\sim$ 20\% was found in the the galaxy centre, consistent with the detection of a broad component in H$\alpha$. Besides, the maps of mean age and mean Z show a trend consistent with the stellar population becoming more metal poor and younger with increasing distance from the galaxy center.

\item The analysis of the emission line intensities revels that in the inner kpc of the NGC~5044, the gas distribution is irregular and the H$\alpha$ emission follows the extended filaments. The [NII] and [SII] emission is concentrated in the nuclear region, while the [OIII] emission is detected over the whole FoV. 

\item To fit the central region H$\alpha$ emission lines a broad component is required, with FWHM $\sim$ 3000\,\kms\, confirming that the featureless continuum emission detected in the stellar population fits indeed is due to a low luminosity active galactic nuclei. 

\item The FWHM of H$\alpha$ narrow component present strong variations over the FoV, from FWHM $\sim$ 150\,\kms\ at the outer regions to 1080\,\kms\ in the galaxy centre. 

\item The dominant excitation mechanism in the filaments is compatible with shock waves with velocity 550\,\kms\, while in the centre it is explained by a LLAGN. Furthermore, we found that the brighter peak located at east (region 1) and at the south (region 3) in the channel maps are co-spatial with the CO emission detected by ALMA. 

\item From the H$\alpha$ velocity field map, we found evidence of rotating gas.  Assuming Keplerian rotation we estimated a mass of 1.9$\times10^{9}M_{\odot}$ within 136~pc. Beside, using $\sigma_\star$ of 279 \kms and M-$\sigma$ relation the estimated $M_{bh}$ found is $1.8\pm1.6\times10^{9}\,M_{\odot}$ . 

\item Filamentary structures, with non-circular motions around $\mid$40$\mid$~\kms are revelled in the residual velocity field map, obtained subtracting the disk model
from the observed radial velocity field. By using 3D model for these structures  we conclude that the gas inside them is falling towards the nucleus.
The inflow rate of  0.4 $M_\odot\,yr^{-1}$, was derived from H$\alpha$ luminosity of $2.7\times10^{41}$
erg/s and with the mean inflow velocity, $\sim$90\,km/s,  estimated for the 3D model. We point out that the inflow rate is $10^2$ larger  than  the accretion rate
necessary to power the LLAGN.

}

\end{itemize}

\section*{Acknowledgements}

{ We thank the anonymous referee for his/her careful reading of our manuscript and for his/her insightful comments
and suggestions that helped to improve the manuscript.} J. A. H.J. thanks to Brazilian  institution CNPq for financial support through  postdoctoral fellowships (projects 158962/2014-1 and 150237/2017-0). R.R. Thanks to CNPq and FAPERGS for partial financial support to this project. R. A. R. thanks financial support from Brazilian institution CNPq (project 303373/2016-4). T.V.R. acknowledges CNPq for the financial support under the grant 304321/2016-8. Based on observations obtained at the Gemini Observatory, which is operated by the Association of Universities for Research in Astronomy, Inc., under a cooperative agreement with the NSF on behalf of the Gemini partnership: the National Science Foundation (United States), the National Research Council (Canada), CONICYT (Chile), Ministerio de Ciencia, Tecnolog\'{i}a e Innovaci\'{o}n Productiva (Argentina), and Minist\'{e}rio da Ci\^{e}ncia, Tecnologia e Inova\c{c}\~{a}o (Brazil).

\end{document}